\documentclass[journal]{IEEEtran}

\usepackage{cite}
\usepackage{amsmath,amssymb,amsfonts}
\usepackage{graphicx,color}
\usepackage{textcomp}
\usepackage{xcolor}
\usepackage{hyperref}
\usepackage{algpseudocode}
\usepackage[ruled,vlined]{algorithm2e}
\usepackage{caption}
\usepackage{subcaption}
\usepackage{makecell}

\newcolumntype{L}[1]{>{\raggedright\let\newline\\\arraybackslash\hspace{0pt}}m{#1}}
\newcolumntype{C}[1]{>{\centering\let\newline\\\arraybackslash\hspace{0pt}}m{#1}}
\newcolumntype{R}[1]{>{\raggedleft\let\newline\\\arraybackslash\hspace{0pt}}m{#1}}

\algnewcommand\algorithmicforeach{\textbf{for each}}
\algdef{S}[FOR]{ForEach}[1]{\algorithmicforeach\ #1\ \algorithmicdo}

\def\BibTeX{{\rm B\kern-.05em{\sc i\kern-.025em b}\kern-.08em
    T\kern-.1667em\lower.7ex\hbox{E}\kern-.125emX}}
\AtBeginDocument{\definecolor{tmlcncolor}{cmyk}{0.93,0.59,0.15,0.02}\definecolor{NavyBlue}{RGB}{0,86,125}}
\usepackage{array}
\usepackage{tablefootnote}
\usepackage[flushleft]{threeparttable}
\usepackage{xfrac}

\begin{document}
%
\title{UCINet0: A Machine Learning based Receiver for 5G NR PUCCH Format 0}
%
%
%

\author{Anil~Kumar~Yerrapragada,~\IEEEmembership{Member,~IEEE,}
        Jeeva~Keshav~Sattianarayanin,~\IEEEmembership{Student~Member,~IEEE,}
        and~Radha~Krishna~Ganti,~\IEEEmembership{Member,~IEEE,}

}
\author{\IEEEauthorblockN{Jeeva~Keshav~Sattianarayanin\IEEEauthorrefmark{1},  Anil~Kumar~Yerrapragada\IEEEauthorrefmark{2},
Radha~Krishna~Ganti\IEEEauthorrefmark{3}}\\
\IEEEauthorblockA{ Department of Electrical Engineering,
Indian Institute of Technology Madras, Chennai, India, 600036 \\
Email: \IEEEauthorrefmark{1}jeevakeshavs@smail.iitm.ac.in, \IEEEauthorrefmark{2}anilkumar@5gtbiitm.in,
\IEEEauthorrefmark{3}rganti@ee.iitm.ac.in
}
\thanks{The first two authors contributed equally to this work. This work was supported by the Indian Ministry of Electronics and Information Technology (MeitY) and the Department of Telecommunications (DoT) under the project ``Next Generation Wireless Research and Standardization on 5G and Beyond" with grant number SP21221155EEMEIT008073 through the 5G Testbed Project.}
}

\maketitle

\begin{abstract}
Accurate decoding of Uplink Control Information (UCI) on the Physical Uplink Control Channel (PUCCH) is essential for enabling 5G wireless links. This paper explores an AI/ML-based receiver design for PUCCH Format 0. Format 0 signaling encodes the UCI content within the phase of a known base waveform and even supports multiplexing of up to 12 users within the same time-frequency resources. The proposed neural network classifier, which we term UCINet0, is capable of predicting when no user is transmitting on the PUCCH, as well as decoding the UCI content for any number of multiplexed users (up to 12). {The test results with simulated, hardware-captured (lab) and field datasets show that the UCINet0 model outperforms conventional correlation-based decoders across all SNR ranges and multiple fading scenarios.}
\end{abstract}



\maketitle

\section{INTRODUCTION}

\IEEEPARstart{T}{he} successful establishment of any wireless communication link between two entities relies on feedback signalling from both ends to indicate the channel quality as well as the status of previous transmissions. The Third Generation Partnership Project (3GPP) standards define the 5G New Radio (NR) Physical Uplink Control Channel (PUCCH)~\cite{5g_bullets}, which is the key enabler of such feedback in the uplink direction. It is a dedicated channel on which a User Equipment (UE) can send control information to a Base station (gNB). Uplink Control Information (UCI) carried by the PUCCH includes (1) Hybrid Automatic Repeat Request (HARQ) acknowledgements for prior downlink transmissions (gNB to UE), (2) Scheduling Request (SR) for the subsequent allocation of uplink transmission resources, and (3) Downlink Channel State Information (CSI) reports containing channel quality metrics that facilitate link adaptation, precoding, and downlink resource allocation. 

In order to encompass the wide range of capacity and latency requirements of various 5G applications, the 5G standards~\cite{3gpp_38_211, 3gpp_38_212, 3gpp_38_213} have provisioned five different formats of the PUCCH transmission. See Table~\ref{tab: pucch_summary} for a summary of the five different formats. Smaller UCI payloads of 1 or 2 HARQ bits and/or an SR are transferred using Formats 0 and 1. Larger payloads, including several SR and HARQ bits, along with CSI reports, are transmitted using Formats 2, 3, and 4. Furthermore, Formats 0 and 2 are suitable for low latency applications since they occupy only 1 or 2 OFDM symbols in the time domain. Formats 1, 3, and 4 are suitable for applications requiring improved coverage and capacity since they extend between 4 and 14 OFDM symbols in the time domain and can provide SNR gain. 

\begin{table*}[h]
\begin{threeparttable}
    \caption{Summary of 5G NR PUCCH formats}
    \centering
    \begin{tabular}{|C{0.8cm}|C{2.3cm}|C{1.1cm}|C{1.2cm}|C{1.2cm}|C{1.9cm}|C{2.2cm}|C{1.2cm}|C{1.45cm}|}
        \hline
        \textbf{Format} & \textbf{Payload} & \textbf{Duration} & \textbf{Duration in symbols} & \textbf{Resource Blocks} & \textbf{Waveform} & \textbf{Modulation} & \textbf{DMRS} & \textbf{Multiplexing} \\ 
        \hline
        \hline
        \textbf{0} & HARQ, SR & Short & 1-2 & 1 & CP-OFDM & None (Sequence-based) & No & Yes\\
        \hline
        \textbf{1} & HARQ, SR & Long & 4-14 & 1 & CP-OFDM & BPSK or QPSK & Yes & Yes\\
        \hline
        \textbf{2} & HARQ, SR, CSI & Short & 1-2 & 1-16 & CP-OFDM & QPSK & Yes & No\\
        \hline
        \textbf{3} & HARQ, SR, CSI & Long & 4-14 & 1-16\tnote{1} & DFT-S-OFDM & $\frac{\pi}{2}$-BPSK or QPSK & Yes & No\\
        \hline
        \textbf{4} & HARQ, SR, CSI  & Long & 4-14 & 1 & DFT-S-OFDM & $\frac{\pi}{2}$-BPSK or QPSK & Yes & Yes\\
        \hline
    \end{tabular}
    \begin{tablenotes}
    \item[1] Limited to 1-6, 8-10, 12, 15, 16 for ease of DFT-S-OFDM.
  \end{tablenotes}
    \label{tab: pucch_summary}
\end{threeparttable}
\end{table*}

In this paper, we focus on  PUCCH Format 0, a particularly important format first used during several steps of the 5G initial attach process~\cite{5g_bullets} and for subsequent feedback transmissions (small payload) even after the establishment of the communication link. Specifically, we look at the design and performance analysis of a receiver for Format 0 using AI/ML techniques such as Neural Network (NN) classifiers. In this paper, we make the following contributions: 

\begin{itemize}
    \item PUCCH Format 0 uses a waveform-based technique for information transfer. It encodes the UCI within phase rotations applied to a pre-defined base sequence. This paper presents a Machine Learning based decoder for Format 0 that uses a neural network to predict the phase rotation. This paper, an extension of our previous work~\cite{yerrapragada2023machine}, is one of the early attempts at designing such a receiver for 5G PUCCH Format 0. 
    \item The integration of AI capabilities with communication systems can be done by replacing existing functionalities with AI-based approaches, by adding new AI-based functionalities, or by adding AI assistance to existing approaches. This paper proposes that the decoding of Format 0 could be performed by an AI/ML model, in lieu of conventional approaches such as correlation with the known base sequences. The AI/ML model takes only the extracted Format 0 resources from the Resource Grid as input and the functioning of this model is independent of the rest of the receiver chain. 
    \item Format 0 signals can be transmitted either by a single UE or by multiple UEs (a maximum of 12 UEs~\cite{3gpp_38_211}) multiplexed in the same time-frequency resources. The AI/ML model described in this paper is capable of decoding the UCI from a single UE up to a maximum of 12 multiplexed UEs, with each UE transmitting any of the allowed combinations of HARQ and SR payloads.
    \item One of the challenges of correlation-based methods is their susceptibility to false detections. Conventional approaches to determine whether a UE has actually transmitted on the PUCCH involve signal power-based thresholds, which are often determined either based on observation or tedious field measurements. We show that the UCINet0 model described in this paper is robust enough to identify false transmissions, thus eliminating the need for thresholds.
    \item This paper provides insights into the generation of training and testing wireless signal datasets derived from both MATLAB simulations and real-time over-the-air captures taken at our indigenous 5G Testbed at IIT Madras~\cite{5gtbiitm}. 
    \item The proposed AI/ML model relies on extracting a map between the received PUCCH samples (input) and the applied phase rotation to the base sequence (output). At least two degrees of randomness are embedded in the input data. The first comes from the phase rotation, which is a function of the random UCI bits. The second comes from the fading channel and thermal noise, which we assume to be Additive White Gaussian Noise (AWGN). In the presence of such randomness, neural networks often converge to the correct input-output relationship while obscuring the path they take to achieve convergence. This paper attempts to offer some insights into the interpretability of the model's behavior.
    \item AI/ML models must adhere to certain complexity, memory, and latency constraints to become deployable on hardware devices such as Field Programmable Gate Arrays (FPGAs). To this end, this paper provides a complexity analysis of the NN models presented. 
\end{itemize}
\section{BACKGROUND ON 5G NR PUCCH FORMAT 0} \label{sec: background_pucch_f0}
PUCCH Format 0 is used to transfer one or two HARQ acknowledgments and/or an SR~\cite{3gpp_38_213}. In 5G, the use of Format 0 signalling begins as early as the initial attach procedure in which a UE is trying to latch onto a gNB. During the initial attach, there are several critical downlink transmissions, such as the Radio Resource Control (RRC) Setup, that need to be acknowledged, which the UE signals using PUCCH Format 0. It is to be noted that at this early stage of the connection process, neither the UE nor the gNB is expected to have obtained detailed information about the wireless surroundings. Consequently, Format 0 employs a sequence-based transmission in which phase-rotated versions of a pre-defined base sequence are transmitted by the UE. The UCI is encoded in the phase of the base sequence. Since the base sequence is assumed to be known at both ends, Format 0 signalling does not contain any pilot reference signals such as the Demodulation Reference Signals (DMRS), nor does it use Quadrature Amplitude Modulation (QAM) to modulate the UCI bits~\cite{kundu2018physical}.  

\subsection{TIME-FREQUENCY ALLOCATION, SEQUENCE GENERATION and UCI ENCODING}
The cyclically shifted base sequence is mapped to the Resource Grid, where it occupies one Resource Block (RB) in the frequency domain and either one or two symbols in the time domain. Note that, one Resource Block in the frequency domain contains 12 Resource Elements (or 12 sub-carriers of the OFDM grid i.e., $N_{sc}^{RB}=12$). Since PUCCH Format 0 can span two symbols, either 12 or 24 Resource Elements can be occupied. The second symbol contains a sequence similar to that of the first symbol and can be used for SNR enhancement at the receiver. In more concrete terms, the PUCCH Format 0 sequence in the frequency domain includes a phase rotation $\alpha$ applied to a low Peak-to-Average Power Ratio (PAPR) base-sequence~\cite{3gpp_38_211} $r_{u,v}^{\alpha}(k)$, and is given by, 

\begin{equation}
\begin{split}
    r_{u,v}^{\alpha}(k) & = e^{j\alpha k}\cdot\bar{r}_{u,v} (k) \\
                                 & = e^{j\alpha k}\cdot e^{j\phi (k) \pi /4},
\end{split}
\label{eq: rotated_base_seq}
\end{equation}
where $k = 0,1,2,..., N_{sc}^{RB}-1$ and $\phi(k)$ is given by Table 5.2.2.2-2 in~\cite{3gpp_38_211}, which is a set of phase factors that lead to the generation of low PAPR sequences. The subscripts $u$ and $v$ represent the group number and the sequence number within the group, respectively. The values of $u$ and $v$ are configured by higher layer group hopping parameters. We note that when PUCCH Format 0 is two OFDM symbols long, intra-slot frequency hopping can be enabled. In this paper, for ease of exposition, we assume that intra-slot frequency hopping is disabled and utilize only one symbol for decoding the UCI bits.

The cyclic shift $\alpha$ applied to the base sequence is given by,
\begin{equation}
    \alpha = \frac{2\pi}{N_{sc}^{RB}}\bigg((m_{0} + m_{cs} + n_{cs}(n_{s,f}^{\mu},l+l'))\hspace{-8pt}\mod N_{sc}^{RB} \bigg),
    \label{eq: alpha}
\end{equation}
where 
\begin{itemize}
    \item $m_{0}\in \{0,1,2,\dots,11\}$ is the initial cyclic shift that is used to differentiate various multiplexed users as defined in ~\cite{3gpp_38_211}. 
    \item $m_{cs}$ is the UCI-specific cyclic shift~\cite{3gpp_38_211} shown in Table ~\ref{tab: mcs_table}. 
    \item $n_{cs}(n_{s,f}^{\mu},l+l')$ is a function based on the pseudo-random binary sequence defined in~\cite{3gpp_38_211}.
    \begin{itemize}
    \item $n_{s,f}^{\mu}$ is the slot number in the radio frame~\cite{3gpp_38_211}.
    \item $l$ is the OFDM symbol number in the PUCCH transmission where $l=0$ corresponds to the first OFDM symbol of the PUCCH transmission~\cite{3gpp_38_211, 3gpp_38_213}.
    \item $l'$ is the index of the OFDM symbol in the slot that corresponds to the first OFDM symbol of the PUCCH transmission in the slot~\cite{3gpp_38_211, 3gpp_38_213}.
    \end{itemize}
\end{itemize} 

While $m_{0}$ and $n_{cs}$ are drawn from higher layer (L2) configurations, a UE scheduled to transmit on a particular time-frequency allocation chooses the $m_{cs}$ depending on the UCI content it transfers, as defined in the Table.~\ref{tab: mcs_table}. In the table, NACK and ACK refer to Negative and Positive Acknowledgments, respectively. Positive SR (i.e., +ve SR) indicates an SR transmission, and a Negative SR (i.e., -ve SR) denotes that an SR is not transmitted despite the PUCCH resource allocation (the UE does not require Physical Uplink Shared Channel (PUSCH) allocations).

\begin{table}[h]
    \centering
    \caption{Possible $m_{cs}$ values in PUCCH Format 0}
    \begin{tabular}{|C{2cm}|L{3.5cm}|}
        \hline
         \textbf{UCI Content} & \textbf{Possible $m_{cs}$ values} \\
         \hline
         1 HARQ & 0 (ACK), 6 (NACK) \\
         \hline
         2  HARQ & 0 (NACK, NACK) \newline 3 (NACK, ACK) \newline 6 (ACK, ACK), \newline 9 (ACK, NACK) \\
         \hline
         1 SR & 0 (+ve SR) \\
         \hline
         1 HARQ \newline + \newline 1 SR & 0 (NACK, -ve SR), \newline 3 (NACK, +ve SR) \newline 6 (ACK, -ve SR), \newline 9 (ACK, +ve SR)\\
         \hline
         2 HARQ \newline + \newline 1 SR & 0 (NACK, NACK, -ve SR), \newline 1 (NACK, NACK, +ve SR) \newline 3 (NACK, ACK, -ve SR), \newline 4 (NACK, ACK, +ve SR) \newline 6 (ACK, ACK, -ve SR), \newline 7 (ACK, ACK, +ve SR) \newline 9 (ACK, NACK, -ve SR), \newline 10 (ACK, NACK, +ve SR)\\
         \hline
    \end{tabular}
    \label{tab: mcs_table}
\end{table}

\subsection{PUCCH FORMAT 0 TRANSMISSION SCENARIOS}
\begin{figure}[ht]
    \centering
    \includegraphics[width=0.48\textwidth]{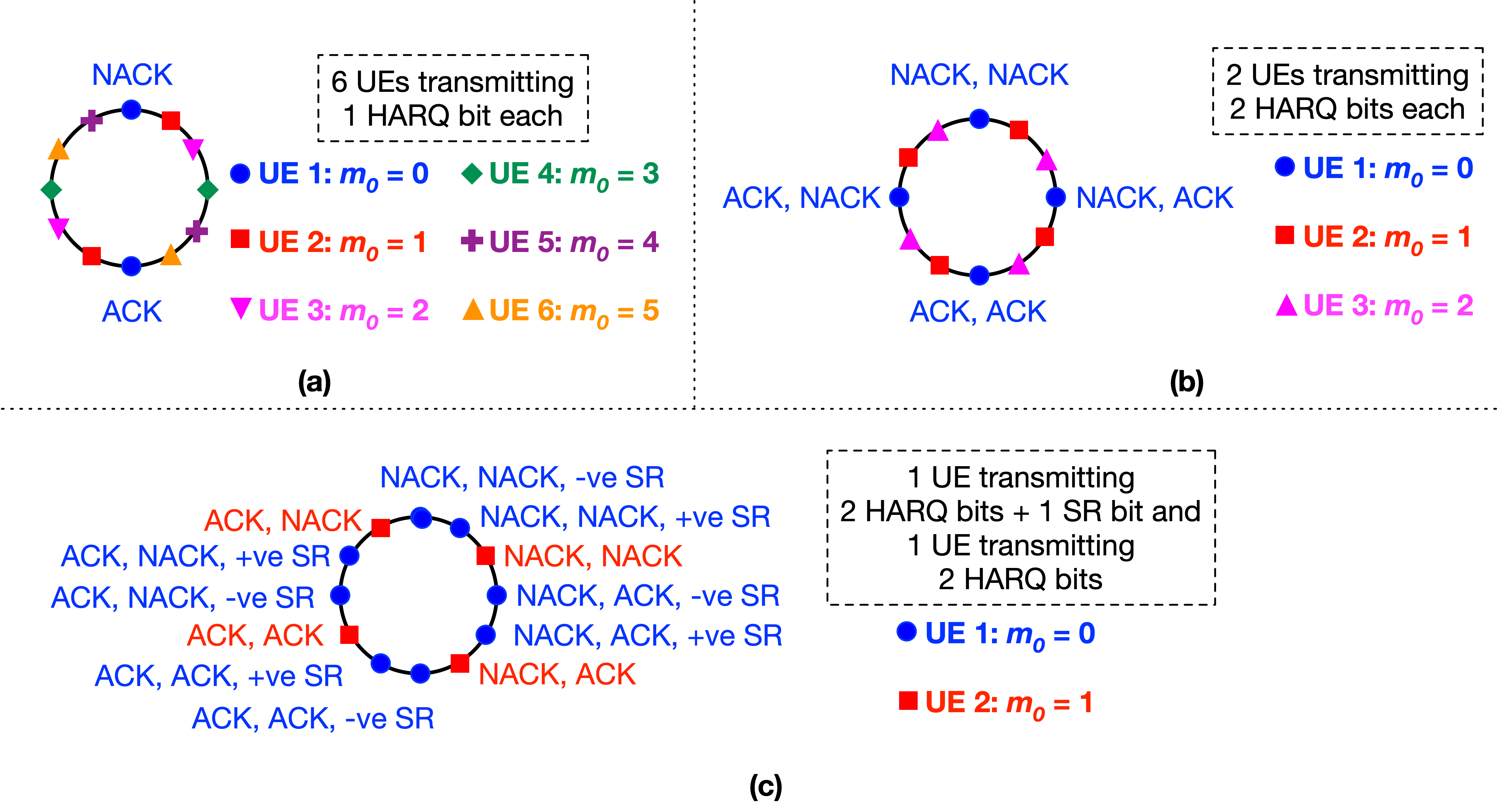}
    \caption{Example scenarios showing how assigning different initial cyclic shifts to different UEs allows them to be multiplexed on the same time-frequency resources.}
    \label{fig: m_cs_mux_ue}
\end{figure}

In the interest of optimizing radio resources, PUCCH Format 0 has a unique potential to multiplex more than one UE on the same time-frequency locations. This is achieved by allocating different initial cyclic shifts ($m_0$) to each UE. Since the number of possible values of $m_0$ is limited to 12, the number of UEs that can be multiplexed is upper-bounded by 12. The maximum number of UEs that can be multiplexed is also driven by the UCI content. A few examples of this are shown in Figure~\ref{fig: m_cs_mux_ue}. Figure~\ref{fig: m_cs_mux_ue}a shows how assigning $m_0$ values of 0 through 5 to a maximum of 6 different UEs allows each of them to pick one of the two possible cyclic shifts and transmit 1 HARQ bit on the same resources. Similarly, Figure~\ref{fig: m_cs_mux_ue}b shows that $m_0$ values of 0 through 2 allow for 3 UEs to be multiplexed when each of them transmits 2 HARQ bits (or 1 HARQ and 1 SR bit) with one of the four possible cyclic shifts.
Additionally, multiplexing across different types of UCI content is possible, wherever applicable. For example, a UE transmitting 2 HARQ bits and an SR bit cannot be multiplexed with another transmitting the same combination, but can be multiplexed with up to 4 more UEs transmitting SR bits or 1 more UE transmitting 2 HARQ bits (Figure~\ref{fig: m_cs_mux_ue}c). Multiplexing of multiple UEs is possible as long as there are no overlaps among the possible cyclic shifts assigned for each UE. 

{\subsection{MATHEMATICAL MODEL OF PUCCH FORMAT 0} \label{subsec: mathematical_model_pucch_f0}
Accounting for UE multiplexing, the generalized equation for the received multi-user Format 0 signal, in the frequency domain is given by,
\begin{equation}
    y(k) = \sum_{m = 0}^{N_{UE}-1} \beta_m h_{m}(k)e^{j\alpha_{m} k}\bar{r}_{u,v} (k) + w(k),
    \label{eq: pucch_rx_signal}
\end{equation}
where $k = 0,1,2,..., N_{sc}^{RB}-1$. Here, $ \beta_m$ indicates whether the $m$-th UE, which was scheduled to transmit on the PUCCH Format 0, transmitted or not, i.e., $\beta_m \in \{0,1\}$. It should be observed that while $N_{UE}$ UEs are scheduled, all of the UEs need not transmit and hence the actual number of transmissions $\tilde{N_{UE}}$ can be less than or equal to $N_{UE}$, i.e., 
\begin{equation}
    \tilde{N}_{UE} = \sum_{m = 0}^{N_{UE}-1} \beta_m \leq N_{UE}.
    \label{beta_m_constraint}
\end{equation}
Let
\begin{equation}
    \Delta = N_{UE} - \tilde{N}_{UE}.
    \label{eq:delta}
\end{equation}
In Eq. \ref{eq: pucch_rx_signal}, $h_{m}(k)$ is the channel between the gNB and the UE transmitting the base sequence with cyclic shift $\alpha_m$ in the $k^{th}$ Resource Element (observe that all multiplexed UEs use the same base sequence). The noise is denoted by $w(k)$.
}

{The goal of PUCCH Format 0 decoding is to determine $\alpha_{m}$ ($m=0, 1, 2, \dots, N_{UE}-1$), given that the gNB observes $y(k)$ ($k = 0,1,2,..., N_{sc}^{RB}-1$). The corresponding optimization problem is shown in the equations below.
}

{
\begin{equation}
    \begin{aligned}
        &\min_{
            \scalebox{0.7}{$\begin{array}{c}
            \overline{\alpha}, \\
            \beta_0, \beta_1, \dots, \beta_{N_{UE}-1}, \\
            \overline{h}_0, \overline{h}_1, \dots, \overline{h}_{N_{UE}-1}
        \end{array}$}}
        \hspace{-2em}
        \sum_{k=0}^{N^{RB}_{sc}-1}\left| y(k) - \sum_{m = 0}^{N_{UE}-1} \beta_{m} h_{m}(k)e^{j\alpha_{m} k}\bar{r}_{u,v} (k)\right|^{2},\\
        &\sum_{m = 0}^{N_{UE}-1} \beta_{m} \leq N_{UE}, \\
        &\overline{\alpha} \in A_0 \times A_1 \times \hdots \times A_{N_{UE}-1}.
    \end{aligned}
    \label{eq: f0_optimization_equation}
\end{equation}
}
{Here, $A_i = \mathcal{R}(\overline{C}_i, {m_0}_i, n_{cs})$ is a set containing $\alpha_{i}$ values for user $i$, given its UCI content vector $\overline{C}_{i}$, its initial cyclic shift $m_{0_i}$ and $n_{cs}$ values. Note that $i = 0,1,\dots, N_{UE}-1$.
}
{For $N_{UE}$ multiplexed users, given their UCI content vectors, their initial cyclic shift and $n_{cs}$ values, the set $\overline{\alpha}$ contains all possible combinations of cyclic shifts $\alpha_{i}$ ($i = 0,1,\dots, N_{UE}-1$). These combinations can be represented as $N_{UE}$-length tuples and the set of all combinations $\overline{\alpha}$ can be obtained by the Cartesian Product of the sets $A_{i}$ where $i = 0,1,\dots, N_{UE}-1$.
}

{Consider the following illustrative examples.} 

{\underline{Example 1:} Assume that $N_{UE} = 1$, $n_{cs} = 0$, $m_{0_0} = 0$ and the UCI content for the scheduled UE is 1 SR and 1 HARQ i.e., $\overline{C}_{0} = [1, 1]$. Then,} 

{\[\overline{\alpha} \in A_0 = \mathcal{R}([1,1],0,0) = \{0, 3, 6, 9\}.\]
}
{\underline{Example 2:} Assume that $N_{UE} = 2$, $n_{cs} = 0$ for both UEs, $m_{0_0} = 0$, $m_{0_1} = 1$ and the UCI content for the first UE is 1 SR and 1 HARQ i.e., $\overline{C}_{0} = [1, 1]$ and that for the second UE is 1 HARQ i.e., $\overline{C}_{1} = [0, 1]$. Then, 
\begin{gather*}
    A_0 = \mathcal{R}([1,1],0,0) = \{0, 3, 6, 9\},   \\
    A_1 = \mathcal{R}([0,1],1,0) = \{1, 7\}, \\
    \overline{\alpha} \in A_0 \times A_1 = \{(0,1), (0,7), (3,1), \\
    (3,7), (6,1), (6,7), (9,1), (9,7)\}.
\end{gather*}
}
{\underline{Example 3:} Assume that $N_{UE} = 3$, $n_{cs} = 0$ for all the three UEs, $m_{0_0} = 0$, $m_{0_1} = 1$, $m_{0_2} = 2$ and the UCI content for the first UE is 1 SR and 1 HARQ i.e., $\overline{C}_{0} = [1, 1]$, that for the second UE is 1 HARQ i.e., $\overline{C}_{1} = [0, 1]$ and that for the third UE is 1 SR i.e., $\overline{C}_{2} = [1, 0]$. Then, 
\begin{gather*}
    A_0 = \mathcal{R}([1,1],0,0) = \{0, 3, 6, 9\},   \\
    A_1 = \mathcal{R}([0,1],1,0) = \{1, 7\}, \\
    A_2 = \mathcal{R}([1,0],2,0) = \{2\}, \\
    \overline{\alpha} \in A_0 \times A_1 \times A_2 = \{(0,1,2), (0,7,2), (3,1,2), \\
    (3,7,2), (6,1,2), (6,7,2), (9,1,2), (9,7,2)\}.
\end{gather*}
}

\subsection{CONVENTIONAL RECEIVER METHODS FOR PUCCH FORMAT 0}

 Since Format 0 PUCCH has no provision for pilots, coherent detection with channel estimation is not possible. Blind correlation methods~\cite{1261943}~\cite{4595664} exist for scenarios in which channel information is unknown, but these methods are often reserved for scenarios in which resource allocation information is also not known. An example of such a scenario is the initial time synchronization between the gNB and the UE. Since the allocation of the synchronization sequences is unknown, longer correlations need to be performed. However, by the time the first Format 0 signal is received at the gNB, time and frequency synchronization is already achieved, and the gNB is aware of the UE's transmission~\cite{wahlqvist1996time}. Since the allocation is known and is always 1 RB long, a simpler length-12 correlation is sufficient.

In ~\cite{kim2020performance, tadavarty2021performance, phan2021enhanced}, various receiver techniques based on correlation are considered. The correlation relies on the fact that the low PAPR base sequence used to encode the UCI content is known to the receiver. The decoding method involves correlating the received samples with various versions of the base sequence that are cyclically shifted. The predicted cyclic shift is the cyclic shift that gives the highest correlation magnitude. It should be noted that peak selection, by definition, requires an optimal threshold to determine whether a peak is due to noise or a true transmission.  

The work in~\cite{kim2020performance} shows that correlation-based signal detection outperforms methods based on raw signal power measurements even in fading channel scenarios. A normalized correlation-based solution for identifying false detections of HARQ bits is provided in~\cite{tadavarty2021performance}. The solution uses the probability distributions of a normalized correlation peak to determine the optimum threshold to classify a reception as false. The receiver algorithm for PUCCH Format 0 proposed in~\cite{phan2021enhanced} eliminates phase opposition across multiple hops by finding the sum of correlation magnitudes across symbols instead of correlation values. 

The common metrics used in the above works are false and missed detections, where false detection refers to the detection of UCI when no UE has transmitted, and missed detection refers to the wrong detection of UCI. We note that the above works identify false transmissions by comparing the correlated peak with a fixed threshold. These fixed thresholds are specific to a certain scenario and are derived from lengthy Monte Carlo simulations. Another significant element that is not considered in~\cite{kim2020performance, tadavarty2021performance, phan2021enhanced} is the multiplexing of multiple UEs in the same resources. This is complicated by the fact that each UE can have different SNRs. It is not clear how the thresholds in the above work extend to multiplexed UEs. Furthermore, fixed thresholds do not account for variations in the received signal power caused by variations in gains of hardware elements such as the LNA and due to varying locations of the multiplexed UEs. For example, a true transmission with large distance-based path loss could be misclassified as false if a fixed threshold is used.

{\subsection{CONVENTIONAL RECEIVER USING DFT-BASED CORRELATION}
As can be seen in Eq.~\eqref{eq: f0_optimization_equation}, the optimal Maximum Likelihood detector is an integer programming optimization problem and does not have a closed-form solution. In particular, solving the optimization problem would require a brute-force search over cyclic shifts $\bar{\alpha}$, transmission indicator $\beta_i$, and channels $h_m(k)$, which is very computationally expensive. A suboptimal receiver is based on the correlation with the base sequence without optimizing over the $\beta_i$ and the channel coefficients.}

{Subsequent to the correlation, the appropriate number of peaks is chosen based on the constraints. In many receivers, including the 5G Testbed deployment at IIT Madras~\cite{5gtbiitm}, the correlation is implemented by an equivalent DFT-based algorithm for decoding the UCI of up to 12 multiplexed UEs. In this paper, we use this method as a baseline for comparison with the performance of the AI/ML model. Comparison with this particular DFT-based method is adequate because, to the best of our knowledge, the other existing academic literature on conventional decoders for Format 0, addresses only the single user scenario, i.e., only one user is transmitting a PUCCH Format 0 signal in a single resource. }
\begin{figure}[h]
    \captionsetup{justification=justified}
     \centering
     \begin{subfigure}[b]{0.48\textwidth}
         \centering
         \includegraphics[width=\textwidth]{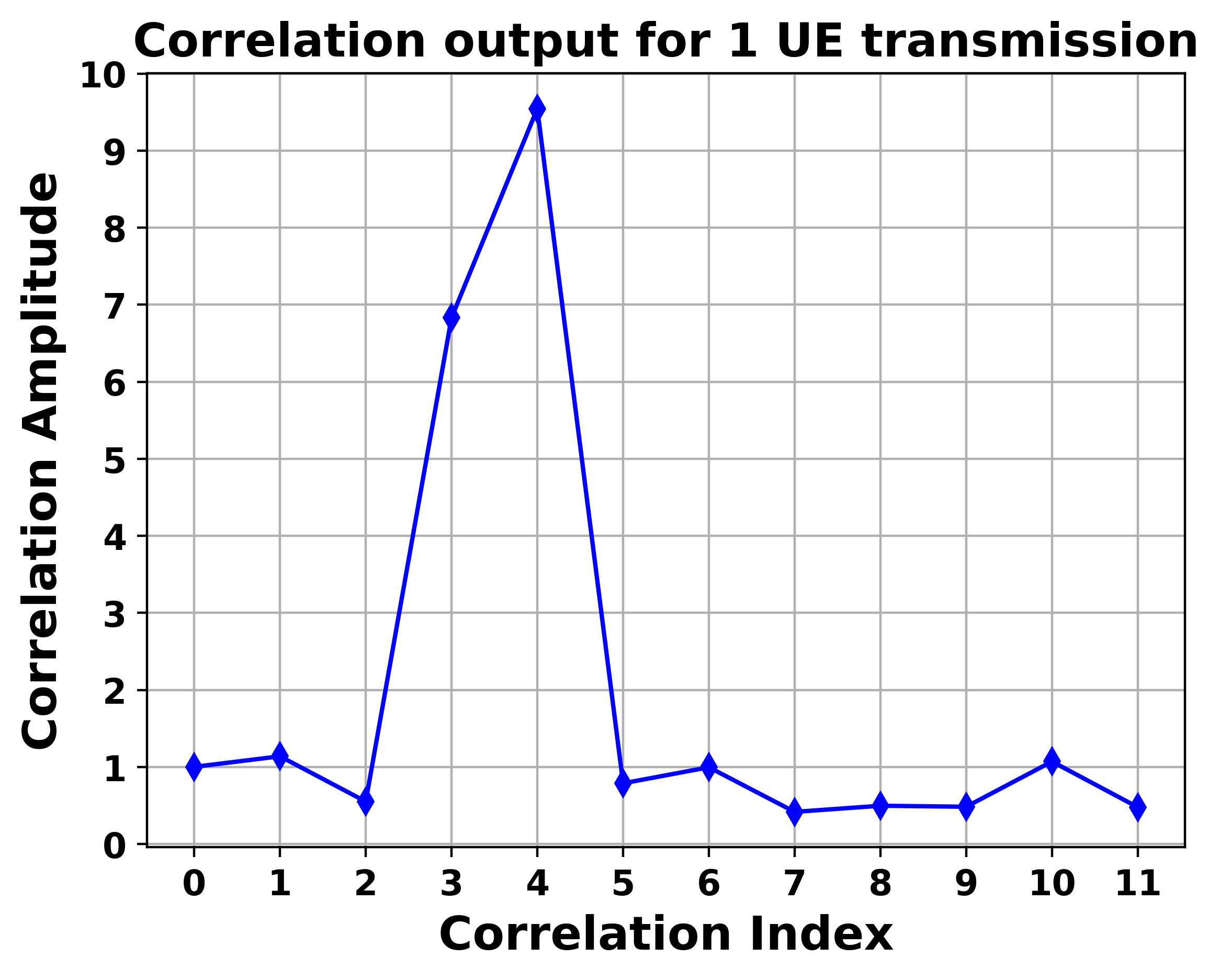}
         \caption{}
         \label{fig: correlation_1_ue}
     \end{subfigure}
     \\
     \begin{subfigure}[b]{0.48\textwidth}
         \centering
         \includegraphics[width=\textwidth]{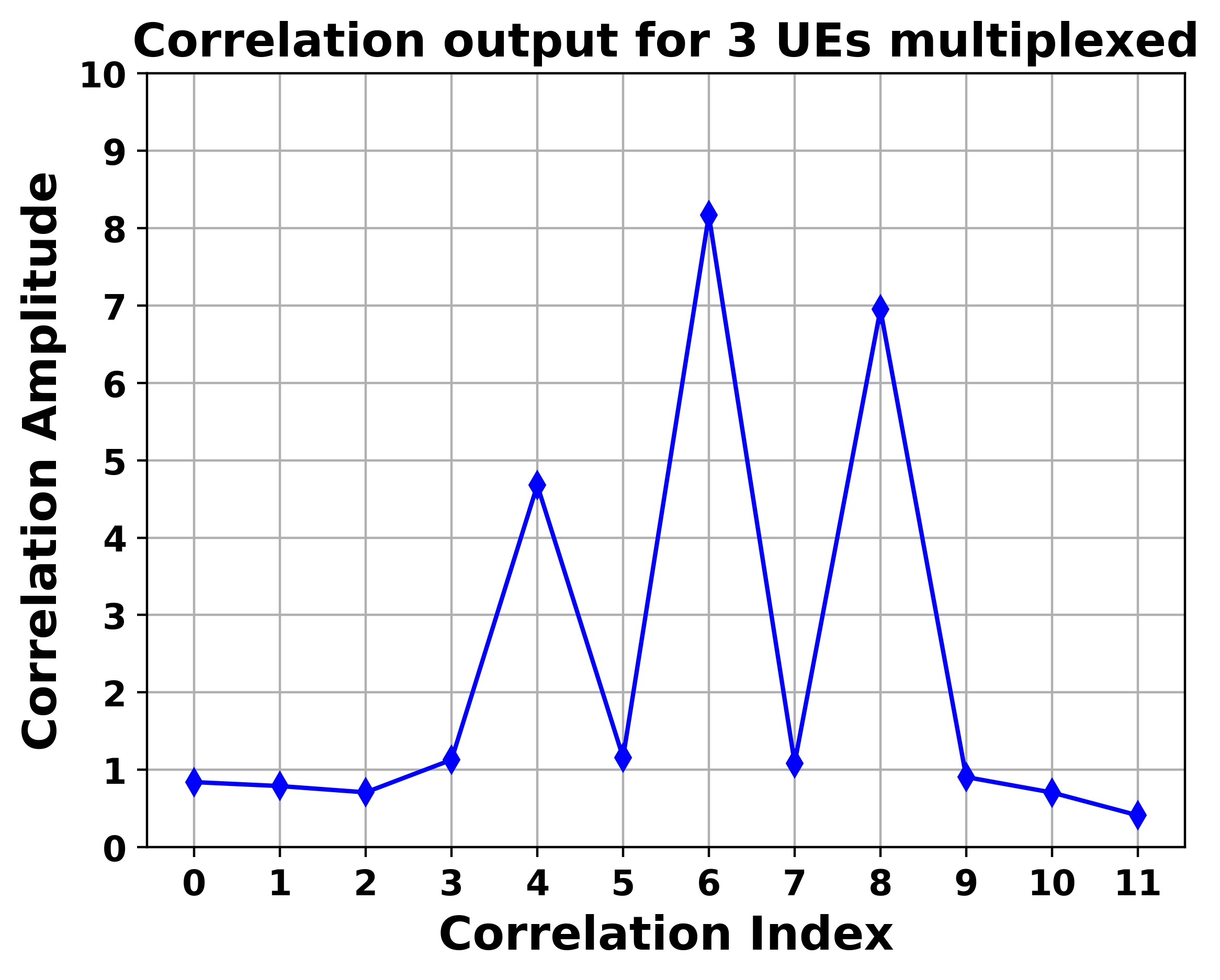}
         \caption{}
         \label{fig: correlation_3_ue}
     \end{subfigure}
        \caption{Illustration of the correlation of the received Format 0 signal with the known base sequence for (a) 1 UE transmission and (b) 3 UEs multiplexed transmission}
        \label{fig: correlation_plots}
\end{figure}
{The first step of the DFT-based correlation is to multiply the received signal by the conjugate of the base sequence to obtain
\[\tilde{y}(k)=y(k).\bar{r}^{*}_{u, v}(k)\stackrel{(a)}{=} \sum_{m = 0}^{N_{UE}-1} \beta_{m} h_{m}(k)e^{j\alpha_{m} k},  k=0\hdots 11, \]
where $(a)$ follows from the fact that $\|\bar{r}_{u,v} (k)\|^2=1$. Note that the noise $w(k)$ is omitted for the sake of being concise. Since the signal spans only one RB (for a $30$ kHz sub carrier spacing, this corresponds to $360$ KHz), a flat fading channel is assumed, \i.e., $h_m(k) = h_m$ for $k=0,1, \dots, 11$. Hence, we obtain}
{\[\tilde{y}(k)=  \sum_{m = 0}^{N_{UE}-1} \beta_{m} h_{m}e^{j\alpha_{m} k}, ~~  k=0,1\hdots, 11. \]}
{We observe that $\tilde{y}(k)$ is a sum of complex exponentials with unknown weights, and the goal is to obtain the frequencies $\alpha_m$ of the different complex exponentials. A 12 point DFT is then applied to $\tilde{y}(k)$ which will peak at the frequencies that are present, i.e., }
{\[ \hat{y}(k)= \text{DFT}(\tilde{y}(k)).\]}
{Based on the constraints sets, the number of users, the location of the peaks of $\hat{y}(k)$ correspond to different $\alpha_{m}$ values. This is illustrated in Figure~\ref{fig: correlation_1_ue} with a correlation peak at $\text{index} = 4$. The correlation index corresponding to the maximum correlation amplitude is taken as the estimated value $\alpha_{m}$. Recall from Eq.~\eqref{eq: alpha} that $\alpha$ is a combination of $m_{0}$, $m_{cs}$ and $n_{cs}$. Therefore, after subtracting $m_{0} + n_{cs}$ from $\alpha$ (subtraction is modulo 12), the cyclic shift specific to UCI $m_{cs}$ remains. In scenarios where multiple users ($N_{UE}$ of them) transmit PUCCH Format 0 signals at the same time-frequency location, the DFT-based correlation algorithm {\em selects the top $N_{UE}$ peaks from the DFT output} and uses them to determine $m_{cs}$ values for each of the $N_{UE}$ multiplexed UEs. Figure~\ref{fig: correlation_3_ue} shows an example in which 3 UEs are multiplexed. In this case, we select the top 3 peaks (at $\text{index} = 4, 6, \text{and } 8$) as $\alpha_m$ values.}

{Observe that the above method is suboptimal, since we do not jointly optimize over $\beta_m$, and the constraint set. Also, the fading is assumed to be flat, which might not be the case with high mobility. However, in real deployments, the above technique is a very practical way of achieving correlation and provides good performance with reasonable implementation complexity.}

\section{MACHINE LEARNING BASED RECEIVER FOR PUCCH FORMAT 0 DETECTION} 
\label{section_ML_F0}
This section lays out the motivation and framework for posing the decoding of PUCCH Format 0 signals as an AI/ML classification problem. Specifically, we show how a single multi-label neural network classifier can serve as a generalized decoder for PUCCH Format 0. 

{The focus of this paper is only on Format 0 since its encoding of information in the phase of a waveform makes it suitable for formulation as an AI/ML classification problem i.e., the prediction of UCI content from a discrete set of phase rotations applied to a base sequence. It should be noted that since Format 0 is sequence based, its decoding does not require any information about the channel. In contrast, other 5G NR PUCCH formats use QPSK modulation of the UCI and require reference signals for channel estimation and equalization. The lengths of the time and frequency allocations are different and variable across the different formats. Other formats also involve scrambling, interleaving, rate matching and error correction mechanisms like Polar Coding and Reed Muller Coding. While it may be possible to bunch Formats 1 to 4 in a multi-task ML problem, Format 0 should be handled separately because of the significant differences in encoding and signal generation.} 


\subsection{MOTIVATION FOR THE APPLICATION OF AI/ML}
Accurate decoding of Format 0 is important because of the following reasons:
\begin{itemize}
    \item Though there are 5 PUCCH formats, in the initial attach procedure, Format 0 is the most feasible Format that can be used to acknowledge and request resources for critical messages to set up a sustained communication link between a UE and a gNB.
    \item Missed detection of a PUCCH Format 0 signal leads to the following scenarios:
    \begin{itemize}
        \item A UE has decoded the gNB’s downlink data and acknowledged it with a HARQ transmission (on PUCCH Format 0 in the uplink direction). If this acknowledgement is missed at the gNB, the whole downlink message will be retransmitted, causing a significant latency in the message delivery and wastage of radio resources.
        \item Another scenario is when a UE’s SR is missed by the gNB. In this case, the UE does not receive an uplink grant and the wait time for the successive grant is usually on the order of 10s of milliseconds. This causes the end-to-end latency of the system to increase.
    \end{itemize}
    \item False detection of a PUCCH Format 0 signal could result in unnecessary uplink grants and, as a consequence, radio resources being blocked.
\end{itemize}

The conventional model described above is based on correlating the received signal with all possible cyclically shifted versions of the base sequence and identifying valid transmissions (This can also be equivalently implemented by the DFT-based approach described above). This is the most common form of implementation of a Maximum Likelihood receiver for Format 0. We encounter the following two issues in designing such a decoder:

\begin{itemize}
    \item In a given time-frequency resource, the number of UEs scheduled to be multiplexed (by Layer 2) is merely an upper bound i.e., if $N_{UE}$ is the scheduled number of UEs, only $\Tilde{N}_{UE}(\leq N_{UE})$ UEs among them might actually transmit. The difference between $N_{UE}$ and $\Tilde{N}_{UE}$ is usually a consequence of one of the following (1) A UE does not transmit any SR despite a PUCCH Format 0 allocation, if no uplink grants are required because of a lack of data in its buffer (2) A UE does not transmit feedback (on HARQ resources) if it does not detect the prior downlink message. Since exact information about $\Tilde{N}_{UE}$ is not known at the receiver, the receiver may falsely detect some transmissions. To mitigate this issue, a threshold could be used to invalidate false transmissions.
    \item Fixed thresholds are hard to design because:
    \begin{itemize}
        \item Based on the proximity to the base station, the correlation peak corresponding to each UE will vary.
        \item In low SNR scenarios, true peaks are not easily distinguishable from noise.
        \item The noise floor of the system can vary with temperature, front-end receiver characteristics, and other hardware-related parameters.
    \end{itemize}
\end{itemize}
The above issues have led us to explore an AI/ML based approach. The data-driven nature of such an approach means that it can learn the hidden patterns from the complex IQ samples extracted directly from the resource grid. 


\begin{table*}[h]
    \centering
    \caption{Differences between prior work \cite{yerrapragada2023machine} and the UCINet0 architecture proposed in this work}
    \begin{tabular}{|C{5 cm}|C{4 cm}|C{5.5 cm}|}
        \hline
         \textbf{Features} & Our prior work\textbf{\cite{yerrapragada2023machine}} & This work (\textbf{UCINet0})\\
         \hline
         Number of multiplexed UEs & 1 & Up to 12 (Max allowed by the spec) \\
         \hline
         Neural Network Input & Rx IQ samples & Rx IQ samples + metadata (number of scheduled UEs) \\
         \hline
         Inclusion of $\Delta$ - the mismatch between the actual number of multiplexed UEs and the upper bound indicated by the L2 scheduler & No & Yes \\
         \hline
          Inclusion of Constraint set mask for inference. See Section~\ref{model_testing} & No & Yes \\
          \hline
         Inclusion of multiple Doppler frequency shifts in the dataset & No & Yes \\
         \hline
         Neural Network output & UCI specific cyclic shift ($m_{cs}$) & Applied cyclic shift to the base sequence ($\alpha$) \\
         \hline
         Neural Network output size & 4 & 12\\
         \hline
         Type of classification & Multi-class (single UE) & Multi-label (Multiple multiplexed UEs) \\
         \hline
         Number of combinations of UCI & 1 (Only 1 SR + 1 HARQ) & 5 (1 SR, 1 HARQ, 2 HARQ, 1 SR+ 1 HARQ, 2 HARQ + SR) \\
         \hline
         Identification of false transmissions & No & Yes \\
         \hline
         Insights into the interpretability of the model & No & Yes \\
         \hline
    \end{tabular}
    \label{tab: differences}
\end{table*}

To our knowledge, there is limited work on applying AI/ML toward the decoding of Format 0 signals. One of the models for Format 0 decoding was proposed in our previous work~\cite{yerrapragada2023machine}. However, this was an initial effort that focused on training the model to predict 1 out of 4 possible $m_{cs}$ values in the specific case of 1 HARQ + 1 SR transmission. The UCINet0 architecture described in this paper extends our previous work to support all the possible UCI payloads shown in Table~\ref{tab: pucch_summary} and all the possible multiplexing combinations of UEs ($N_{UE} = 0, 1, \dots, 12$). Table~\ref{tab: differences} highlights the key differences between our prior work and the generalized architecture described in this paper.


\subsection{PUCCH FORMAT 0 AS AN ML CLASSIFICATION PROBLEM}
Classification, a widespread use case of machine learning algorithms, is a supervised learning task that involves identifying the class (label) to which a given data instance belongs. A predictive model, typically a Neural Network (NN), is trained with a dataset that contains several input instances and the corresponding ground truth class labels. Labeled data helps us ``supervise" the NN in learning the correct input-output mapping. 

Learning happens by iteratively minimizing a loss function, which is a distance metric between the predicted label and the ground truth label. In many classification tasks, the input-output mappings are intractable. ML techniques, such as neural networks, have proven to be more adept at extracting these mappings. 

There are typically three types of classification tasks. The first is Binary Classification, in which the input data falls into one of two classes. An example of this is email classification as spam or not spam. The second is multi-class classification, in which input data could belong to one of more than two classes. A routinely cited example is classifying handwritten digit images into one of ten numbers. The third is multi-label classification, in which input data can independently belong to more than one class. An example of this is the identification of multiple genres a movie belongs to, given a summary of its plot. 

{\underline{Objective of the PUCCH Format 0 Classification Task}: Given a sequence of received frequency domain Format 0 samples, predict the $\Tilde{N}_{UE}$ phase rotation values $\alpha_{0}, \alpha_{1} \dots \alpha_{\Tilde{N}_{UE}-1} $ (see Eq.~\eqref{eq: pucch_rx_signal}) applied to the base sequence. The obtained $\alpha$ values can then be mapped back to the UCI-specific cyclic shift $m_{cs}$ for each UE by modulo 12 subtraction of the corresponding $m_{0}$ and $n_{cs}$, which are provided by the higher layers (L2). 
In the case where no user is transmitting on the PUCCH ($\Tilde{N}_{UE} = 0$), the NN prediction has to reflect this. In summary, the NN classifier has to predict either a single $\alpha$ value, multiple $\alpha$ values, or zero $\alpha$ values. Such a pattern of prediction is a typical use case of multi-label classification. }

\section{NEURAL NETWORK ARCHITECTURE}
\label{section_NN_Arch}
\begin{figure*}[t!]
\centering
\includegraphics[width=0.9\textwidth]{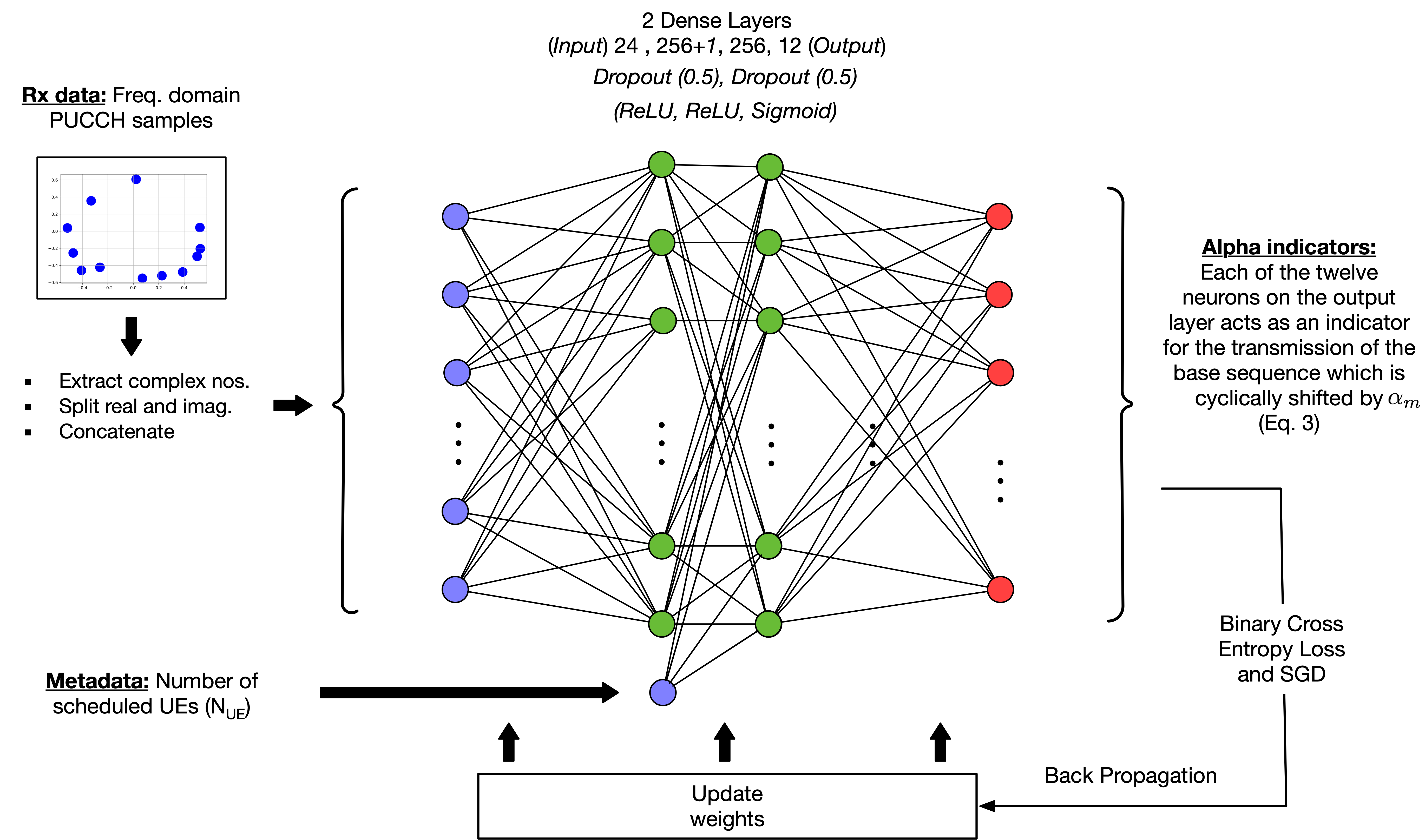}
\caption{{UCINet0 Architecture for PUCCH Format 0 decoding with 24 neurons in the input layer, 256+1 neurons in the second layer, 256 neurons in the third layer, and 12 neurons in the output layer.}}
\label{fig: nn_arch}
\end{figure*}

Given a problem statement, there is no formula for knowing apriori the best neural network architecture. For most problems, initial architectures are determined based on prior experience and domain knowledge. These architectures are then tuned by experiment. Figure~\ref{fig: nn_arch} shows the UCINet0 architecture that showed the best performance in our experiments. In subsequent sections, we also comment on the trade-off between high complexity and performance.

Owing to the relatively small dimension of the input, our proposed UCINet0 is a Fully Connected Neural Network (FCN). {The choice of an FCN also stems from the fact that the phase rotation, containing the UCI, is applied to all elements of the input vector rather than a localized portion (which is what CNNs would be good at extracting). Architectures such as RNNs, LSTM and transformers are not applicable in this case, since we are not dealing with sequential or time series data and there is no need to learn and retain long-term dependencies across the sequence i.e., there is no memory involved.} 

{We also experimented with Complex Valued Neural Networks (CVNNs) in which the model inputs, outputs, and weights corresponding to each neuron are complex. Our ablation studies have shown that the real valued FCNs perform much better than their complex valued counterparts. We concluded that, given the small input size, the relationship between the real and imaginary parts of the input could just as easily be expressed by an equivalent FCN with twice the neurons in the input. However, CVNNs remain an interesting option for future applications of AI/ML in communications.}

{\subsection{ARCHITECTURE DESCRIPTION}
The UCINet0 architecture, as shown in Figure.~\ref{fig: nn_arch} has the following characteristics: }
{\subsubsection{MODEL INPUT}
The input to the NN has two components. The first component is the received PUCCH Format 0 signal. Since the PUCCH Format 0 signals occupy a single resource block, there are $N_{sc}^{RB} = 12$ received complex samples denoted as $\overline{y} \in \mathbb{C}^{N_{sc}^{RB} \times 1}$. We split and concatenate the real and imaginary parts of $\overline{y}$ to obtain $\overline{Y} \in \mathbb{R}^{2N_{sc}^{RB} \times 1}$ as shown below. In this paper, $\overline{Y}$ constitutes the input layer of UCINet0. }
{
\begin{equation}
\begin{split}
    \overline{Y} &= [\text{Re}(\overline{y}), \text{ Im}(\overline{y})]\\
    &= [ \text{Re}(y(0)), \text{ Re}(y(1)), \dots \text{Re}(y(N_{sc}^{RB}-1)), \\
    &\hspace{6mm}\text{Im}(y(0)), \text{ Im}(y(1)), \dots \text{Im}(y(N_{sc}^{RB}-1))]^T.
\end{split}
\label{eq:nn_input}
\end{equation}
}
{The second component of the input, which we feed to the first dense layer, is a metadata to further aid the NN. The metadata indicates the number of UEs scheduled by L2 ($N_{UE}$) in a given time-frequency resource. Even in practical 5G systems, the metadata $N_{UE}$ is known at the L1 (through FAPI protocol). From an AI/ML perspective, it can aid the model to learn information about the number of scheduled users.} 

{However, as mentioned in the previous section, we note that the metadata from L2 need not always match the actual number of transmitting UEs. Hence we defined the quantity $\Delta$ (see \eqref{eq:delta}) to model this mismatch. In this paper, we consider various cases of a maximum metadata offset $\Delta$, which is the difference between the actual number of transmitting UEs and the scheduled number of multiplexed UEs. For example, $\Delta = 0$ indicates that the number of UEs provided as the metadata input to the NN matches the number of UEs actually transmitting in the given allocation. $\Delta = 2$ indicates that for each data instance, the number of UEs provided as the metadata input to the NN could be offset by any value among 0, 1 and 2, with 0 being the best case (no offset) and 2 being the worst. In this paper, during model training, we assume that $\Delta = 2$ and we show inference results for $\Delta = 0, 2, 4$.}

{\subsubsection{HIDDEN LAYERS}
We use 2 hidden layers of size 256. As stated above, the first hidden layer also takes the metadata as an input, in addition to the 256 neurons. Both hidden layers use a Rectified Linear Unit (ReLU) activation function.}
{\subsubsection{MODEL OUTPUT}
The output layer contains twelve neurons. Each output neuron acts as an indicator for the transmission of the base sequence cyclically shifted by $\alpha_m$ (Eq.~\eqref{eq: pucch_rx_signal}), where $m = 0,1,2,\dots, 11$.}

{The output layer uses a Sigmoid activation function. The use of the Sigmoid activation function is motivated by the fact that in a multi-label classifier, the sum of all the output probabilities need not be 1. Each output neuron can be thought of as a part of an independent binary classifier. In other words, since each output neuron corresponds to an $\alpha_m$ value, the probabilities of each output neuron indicate the confidence with which the NN detects the presence of the $m^{th}$ UE.}

\begin{figure}[h!]
    \centering
    \includegraphics[width=0.48\textwidth]{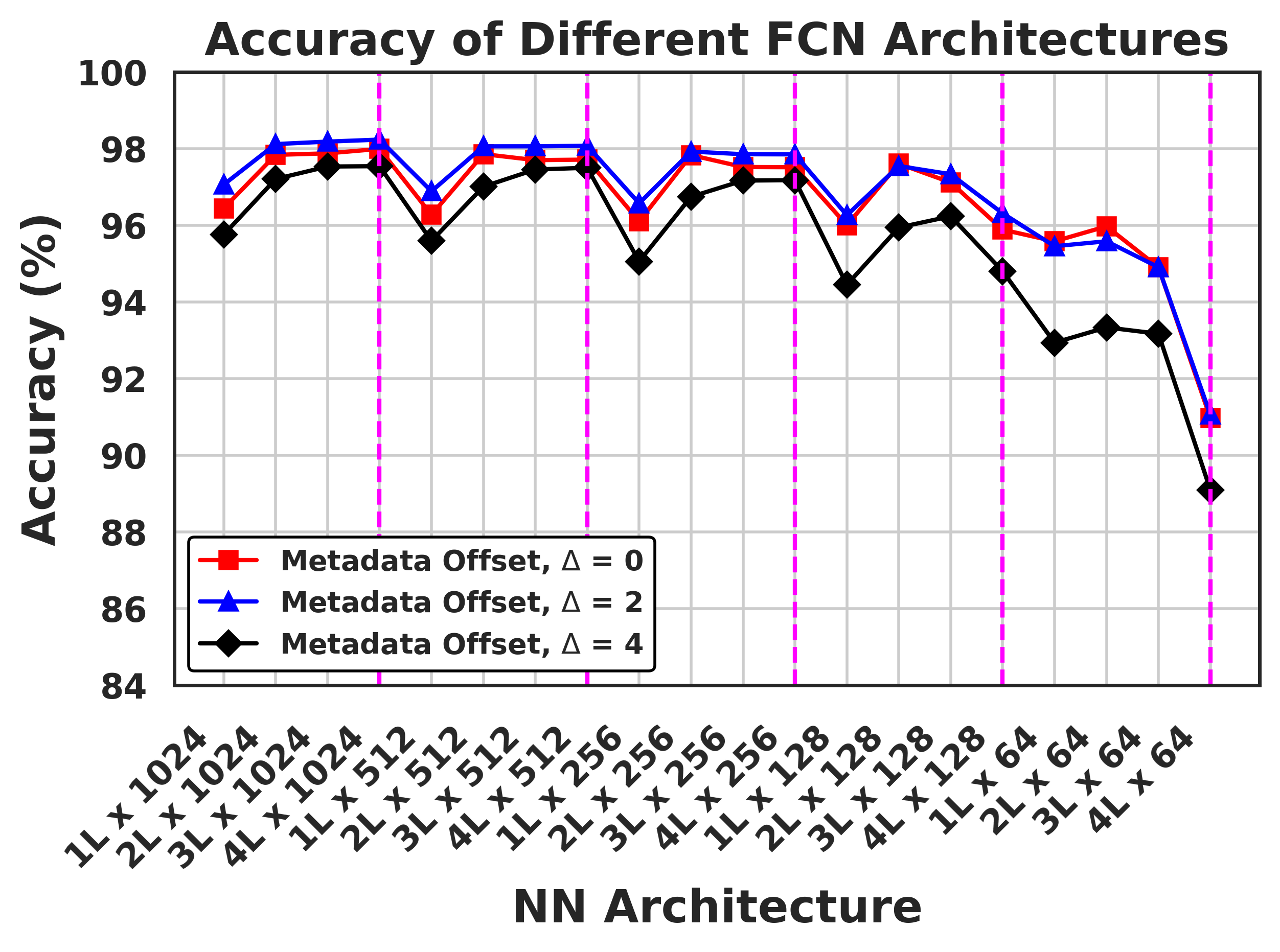}
    \caption{{Model accuracy for different FCN architectures at various values of $\Delta$. Here, the number of FCN layers and the number of neurons of each architecture are presented on the x-axis, and $\Delta$ represents the maximum metadata offset between the true value and L2’s upper bound for the number of multiplexed UEs. We use these results to arrive at the final UCINet0 architecture presented in this paper.}}
    \label{fig: acc_vs_complexity}
\end{figure}

\begin{figure*}[ht!]
    \captionsetup{justification=justified}
     \centering
     \begin{subfigure}[b]{0.48\textwidth}
         \centering
         \includegraphics[width=\textwidth]{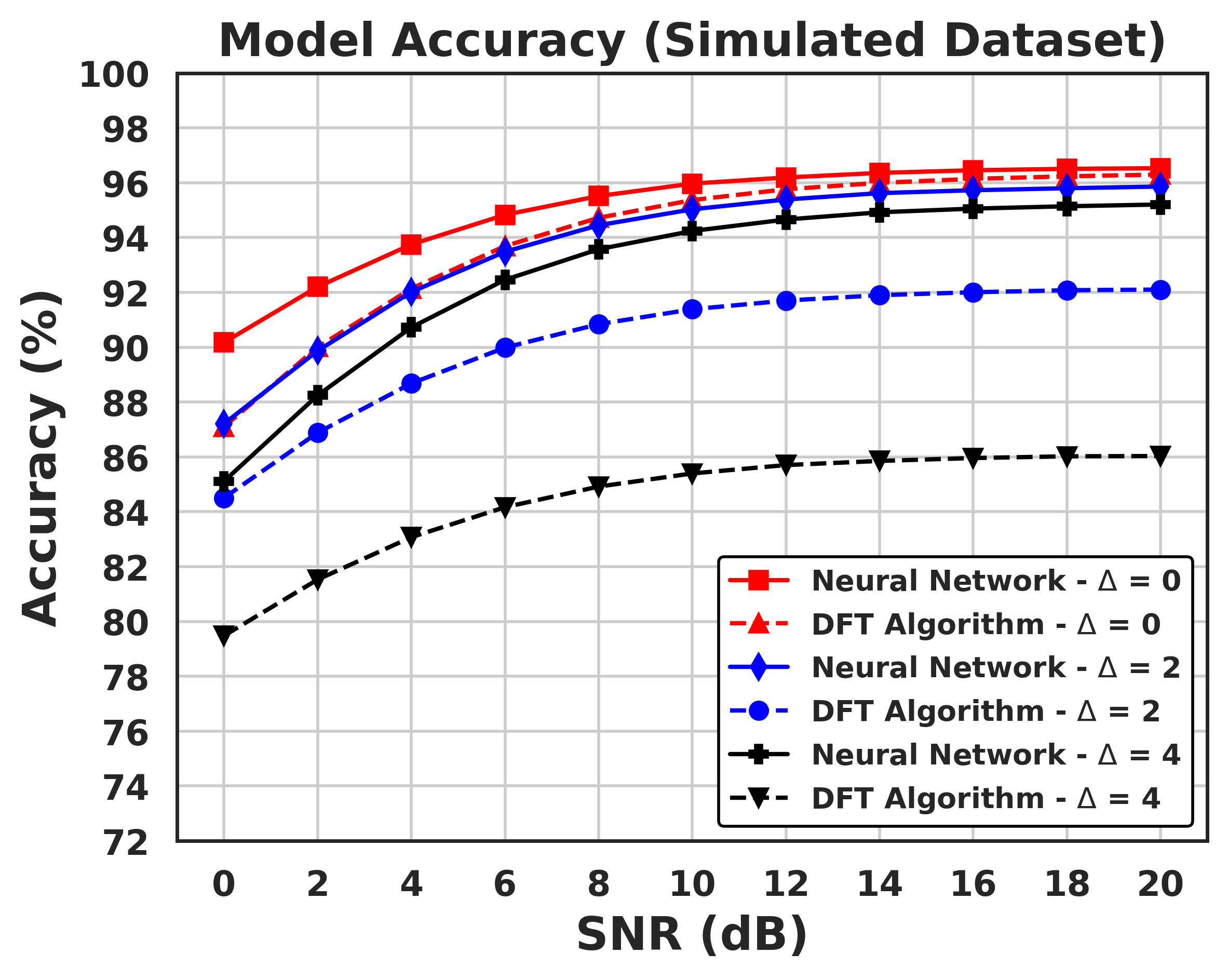}
         \caption{}
         \label{fig: no_metadata}
     \end{subfigure}
     \begin{subfigure}[b]{0.48\textwidth}
         \centering
         \includegraphics[width=\textwidth]{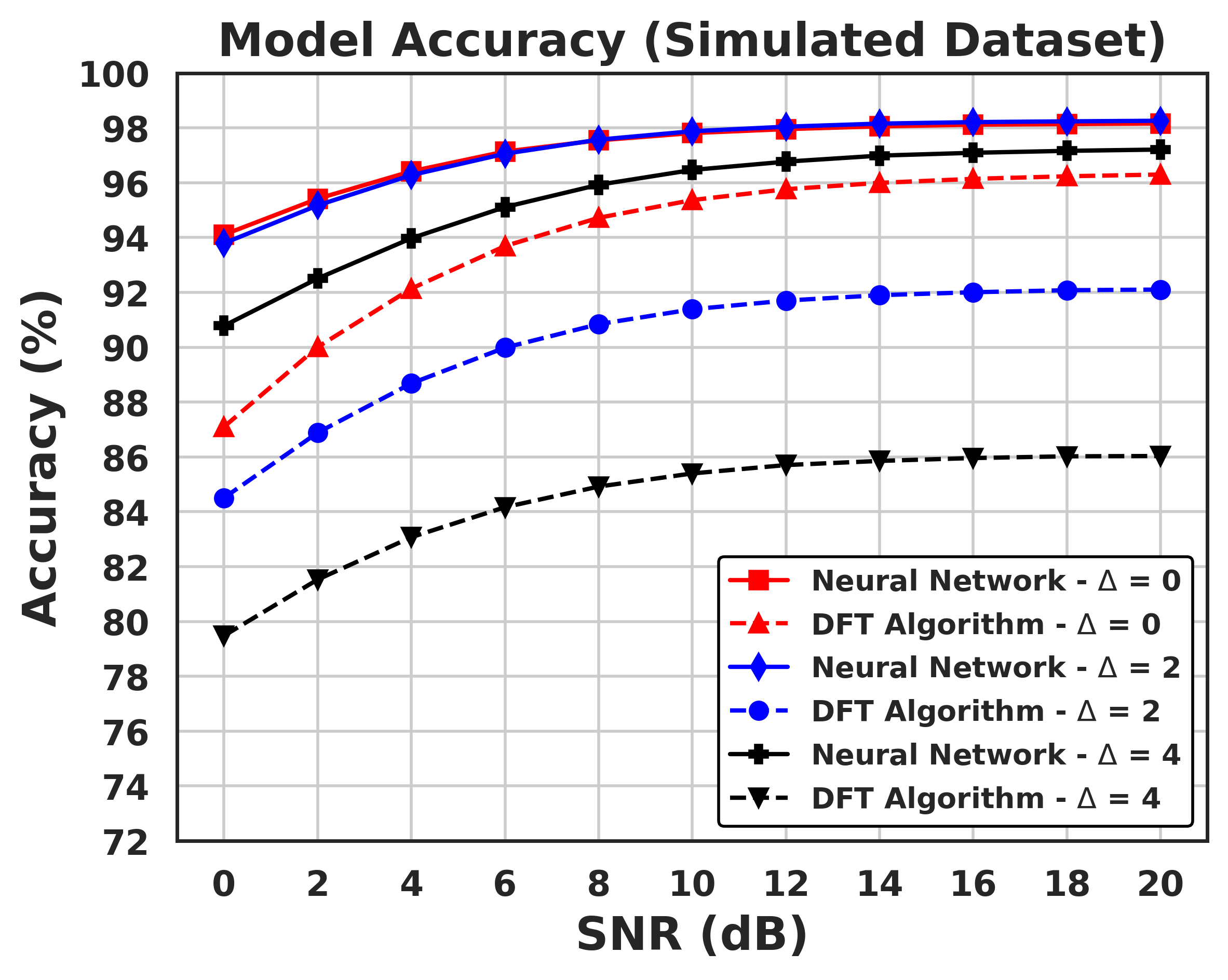}
         \caption{}
         \label{fig: metadata_input_layer}
     \end{subfigure}\\
     \begin{subfigure}[b]{0.48\textwidth}
         \centering
         \includegraphics[width=\textwidth]{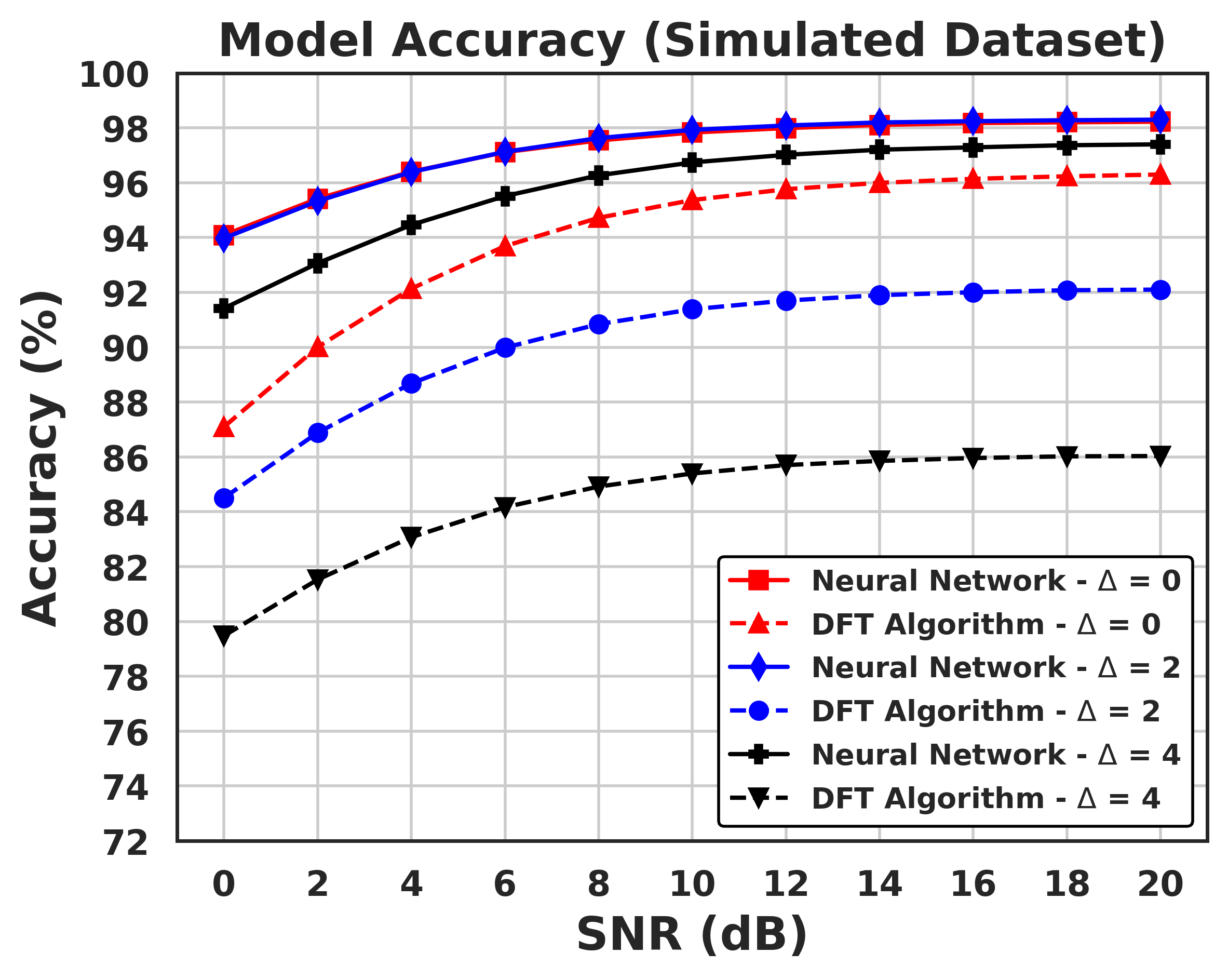}
         \caption{}
         \label{fig: metadata_first_hidden_layer}
     \end{subfigure}
     \begin{subfigure}[b]{0.48\textwidth}
         \centering
         \includegraphics[width=\textwidth]{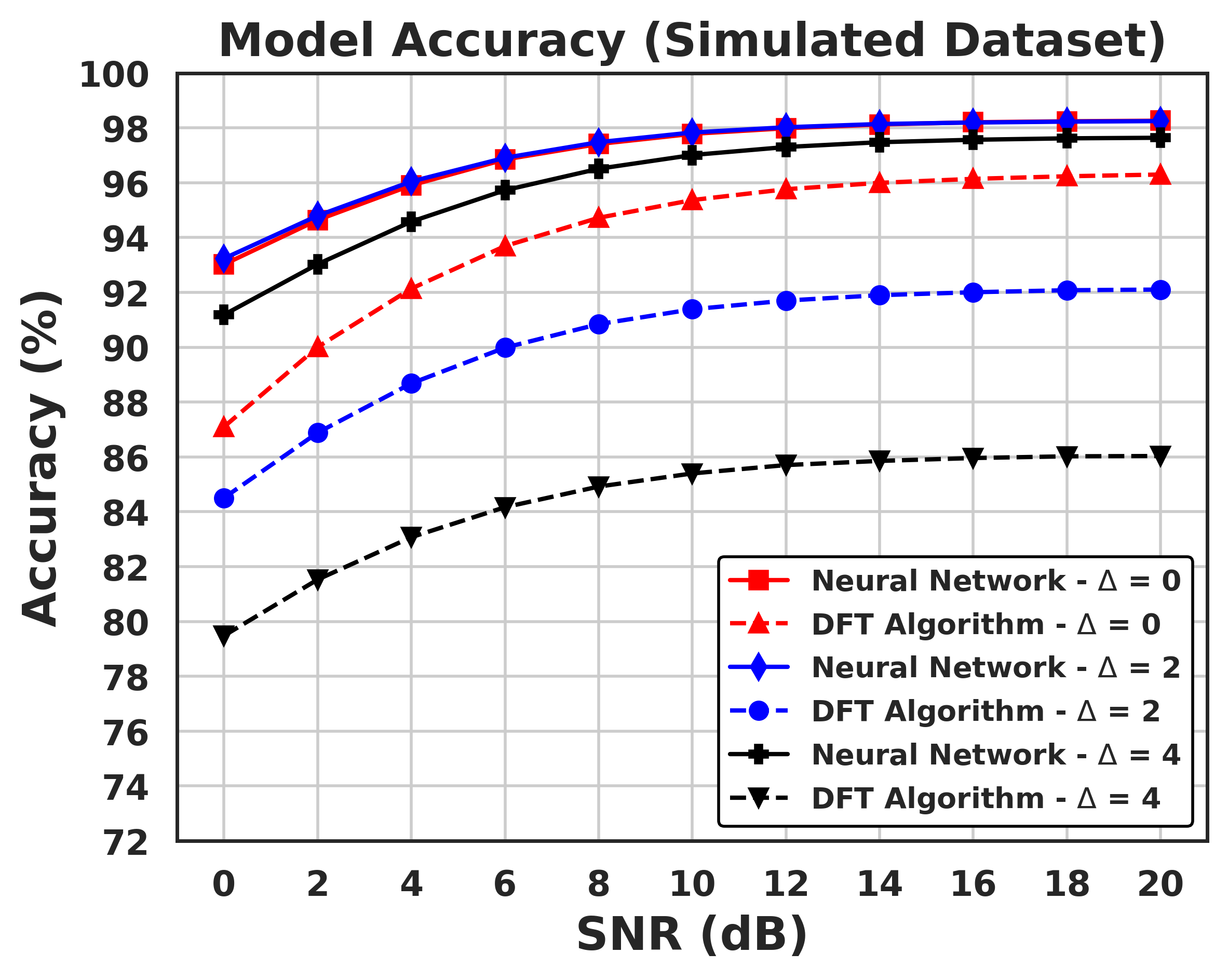}
         \caption{}
         \label{fig: metadata_second_hidden_layer}
     \end{subfigure}
     \caption{{Model test accuracy vs SNR for various values of $\Delta$. (a) No metadata given during training, (b) Metadata given during training at the input layer, (c) Metadata given during training at the first hidden layer, and (d) Metadata given during training at the second hidden layer.}}
     \label{fig: impact_of_metadata}
\end{figure*}

\begin{figure*}[ht!]
    \captionsetup{justification=justified}
     \centering
     \begin{subfigure}[b]{0.48\textwidth}
         \centering
         \includegraphics[width=\textwidth]{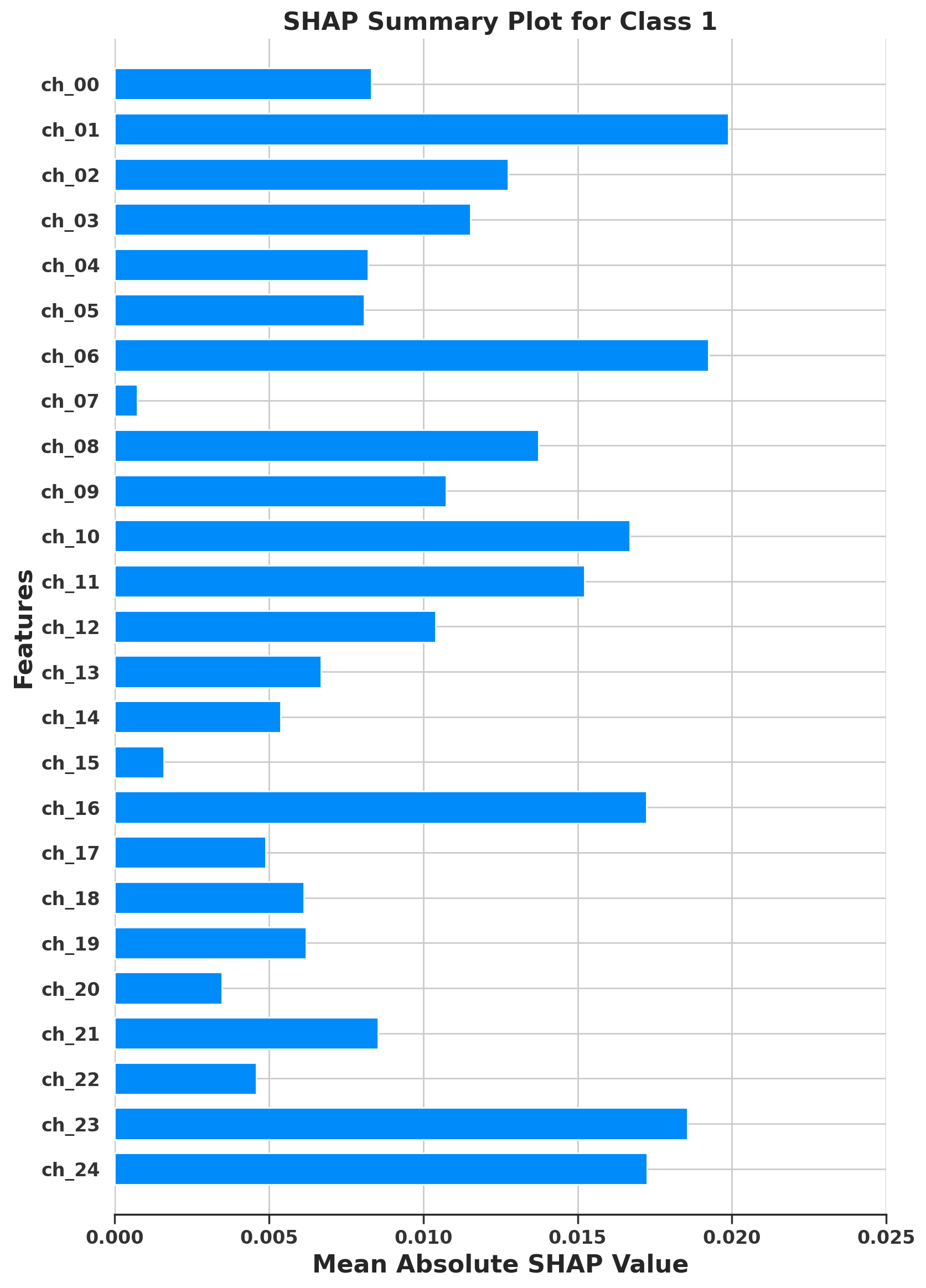}
         \caption{}
         \label{fig: shap_class_x}
     \end{subfigure}
     \begin{subfigure}[b]{0.48\textwidth}
         \centering
         \includegraphics[width=\textwidth]{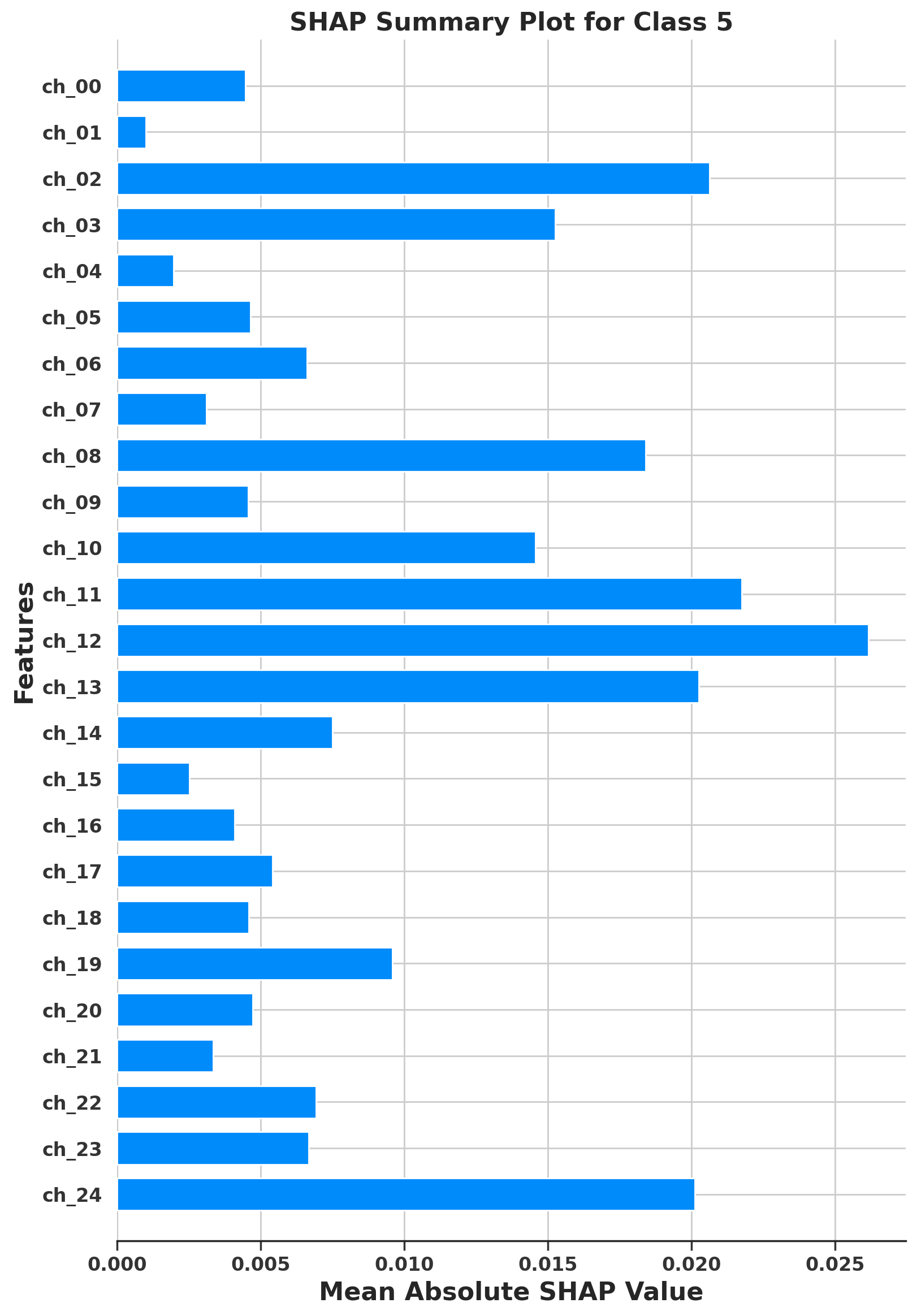}
         \caption{}
         \label{fig: shap_class_y}
     \end{subfigure}
     \caption{{Shapley Values for two example output classes (a) Class 1 (b) Class 5. The distribution of the Shapley values indicates that all input features (both received signal and the metadata) are necessary for the model's prediction.}}
     \label{fig: shap}
\end{figure*}

{\subsection{ABLATION STUDY: FCN ARCHITECTURES}}

{We arrived at the final UCINet0 architecture of 2 layers and 256 neurons after evlauating the performance of other FCN architectures with a varying number of layers and neurons. Figure~\ref{fig: acc_vs_complexity} shows the UCI decoding accuracies (defined in Section~\ref{subsec: accuracy}) obtained with 20 variations of FCN architectures. We considered combinations of $1$, $2$, $3$ and $4$ layer models with $64$, $128$, $256$, $512$ and $1024$ neurons in the hidden layers.}

{From Figure~\ref{fig: acc_vs_complexity}, we observe the following: 
\begin{itemize}
    \item Model performance generally improves as we increase the number of neurons from $64$ to $1024$ for any number of layers and all values of $\Delta$.
    \item For all values of $\Delta$, all of $1024$, $512$, and $256$ neurons result in almost the same accuracy values for a given number of hidden layers. For example, $2 \text{ layers} \times 1024$, $2 \text{ layers} \times 512$ and $2 \text{ layers} \times 256$ result in similar accuracy values.
    \item For all values of $\Delta$, we observe that the accuracy drops initially and then slowly increases as the number of layers increases. This trend becomes more explicit as we move from $1024$ to $128$ neuron models.
\end{itemize} 
}
{Hence, considering the balance between the number of layers, the number of neurons, and accuracies for various values of $\Delta$, the 256 neuron architecture was chosen. Though the $2 \text{ layers} \times 256$ and $3 \text{ layers} \times 256$ architectures exhibit similar performance, the former seemed to be the optimal choice because of its better performance when $\Delta = 0$ and $2$, which are the most common metadata offset scenarios in practical systems. }

{Note that for this ablation study, in each architecture with more than 1 hidden layer, the penultimate hidden layer contains the metadata input as an additional neuron. For architectures with one hidden layer, the metadata is added to the input layer as an additional neuron. The motivation for the inclusion of such metadata is discussed in the subsequent section.}

{
\subsection{ABLATION STUDY: METADATA AS AN INPUT}}
\label{subsubsection_ablation_metadata}
{
To understand the need for metadata in training the neural network, we perform an ablation study of the model (1) with and without metadata, and (2) including metadata at various layers of the model.}

{
Figure~\ref{fig: no_metadata} shows the accuracy of the model across various SNRs when metadata is not fed as input to the neural network. Figure~\ref{fig: metadata_input_layer}-~\ref{fig: metadata_second_hidden_layer} show the accuracy with metadata fed in the input layer, the first dense layer, and the second dense layer, respectively. From Figure~\ref{fig: impact_of_metadata}, it is clear that the inclusion of metadata provides significant gains. It can be observed that the exclusion of metadata leads to a $2\%$ accuracy drop at higher SNRs and $4\%$ accuracy drop at lower SNRs. We observe that the inclusion of metadata at the first hidden layer provides the best accuracy gain. This is reflected very prominently at lower SNRs. Consequently, we reflect this in our UCINet0 model architecture, such that the first hidden layer contains $256+1$ neurons as shown in Figure~\ref{fig: nn_arch}.)}

{
\subsection{SHAPLEY VALUES}
To better understand the impact of various features of the input on the model prediction, we also look at the Shapley values, which explain the contribution of individual features towards a prediction. Figure~\ref {fig: shap}, generated using the SHAP toolbox in Python shows the mean absolute SHAP values for each of the input features, for two example classes ($\alpha = 1$ and $\alpha = 5$). From the figure, we understand the following: (1) the metadata input (ch24) consistently has a high Shapley value, further emphasizing its importance in the model training; (2) among the other input features, i.e., the received PUCCH Format 0 signal, there is no dominant feature. Consider the feature ch01 as an example. In Figure~\ref{fig: shap_class_y}, it has a low SHAP value, indicating its lower contribution for the prediction of Class $5$. However, the same feature has a high SHAP value in the prediction of Class $1$ as shown in Figure~\ref{fig: shap_class_x}. Similar observations can be drawn from other features for different classes, indicating that the model requires all the features to make the best prediction. This conclusion validates our choice of an FCN architecture over CNN.}

{\section{DATA GENERATION}}
Dataset generation in AI/ML for wireless communication has a unique challenge not present in other domains - the non-availability of off-the-shelf benchmark datasets. One way of developing such datasets is through state-of-the-art simulation tools, such as the MATLAB 5G Toolbox, which can generate near-accurate datasets under various channel impairments. Simulated data is a good starting point for training neural networks in communication problems. However, we note that including field data, if available, gives us an insight into the generalization performance of AI/ML models across different distributions of data. The inclusion of field data also aligns with 3GPP requirements for 5G Release 19~\cite{3gpp_38_843}. For this paper, we use a combination of the above two approaches for data generation. The datasets used in this paper are publicly available at~\cite{aiml_pucch_dataset}.

\subsection{SIMULATED DATASETS} 
Using the MATLAB 5G Toolbox, we generate received waveforms containing PUCCH Format 0 signals. These generated waveforms include fading channel impairments and Additive White Gaussian Noise (AWGN). 


The fading channels include Doppler effects corresponding to frequency shifts ($f_{d}$) in the range of $0$ to $2000$ Hz, which map to various UE speeds in the range of $0$ (static) to $600$ kmph (high speed) for a carrier frequency of $3.5$ GHz. These Doppler frequency shifts allow for the simulation of fast-fading channels. For various SNR values in the range of $0$ to $20$ dB, we generate PUCCH signals transmitted over various TDL channels (TDLA30, TDLA300, TDLC30 and TDL300) and store the noisy received samples. These samples are the input to the NN. For each input, the corresponding output label is the applied phase rotation value for each of the $N_{UE}$ UEs in a multi-hot encoded format. The pseudocode for generating PUCCH Format 0 datasets in MATLAB is shown in Algorithm~\ref{alg: data_gen}.


We first place PUCCH Format 0 signals on all 14 OFDM symbols in a slot and 12 resource blocks in each symbol, spaced 20 resource blocks apart. This results in 168 allocations per resource grid. We generate 1000 iterations (indicated by \textit{iter}) of such resource grids. These 1000 grids are generated for each value of $N_{UE}$ in the range 0 to 12. The process is further repeated for the five SNR values. The size of the simulated dataset thus generated is $168000$ per $N_{UE}$ per SNR.

\begin{algorithm}[h]
    \caption{Simulated dataset generation}
    \label{alg: data_gen}
    $\operatorname{\Delta} = \{0, 2, 4\}$ \\
    $\operatorname{N_{UE}} = \{0, 1, 2, \dots, 12\}$ \\
    $\operatorname{SNR} = \{0, 2, 4, 6, 8, 10, 12, 14, 16, 18, 20\}$ dB \\
    $\operatorname{fd} = \{0, 0.4, 0.5, 0.8, 1, 1.2, 1.5, 1.6, 2\}$ kHz \\
    $\operatorname{Iter} = \{1, 2, \dots, 1000$\} \\
    \ForEach{$\operatorname{\Delta}$, $\operatorname{N_{UE}}$, $\operatorname{SNR}$, $\operatorname{fd}$, $\operatorname{Iter}$}{
        $\operatorname{n_{UE}} = 0, 1, 2, \dots, \operatorname{N_{UE}}$ \\
        \ForEach{$u \in \operatorname{n_{UE}}$}{
            \ForEach{$c \in \{1, 2, \dots, 168\}$}{
                bit\_len\_harq $\in \{0, 1, 2\}$ \\
                bit\_len\_sr $\in \{0, 1\}$ \\
                Generate UCI bits and label ($\alpha$) \\
                Generate a Format 0 signal and place it in the resource grid\\
                }
            OFDM Modulation of the TX OFDM grid\\
            Transmit over the channel
        }
        Receive combined signal from $\operatorname{N_{UE}}$ UEs (Eq.~\ref{eq: pucch_rx_signal})\\
        Perform OFDM Demodulation to obtain the RX grid\\
        \ForEach{$c \in \{1, 2, \dots, 168\}$}{
            Extract and save Format 0 signal \\
        }        
    }
\end{algorithm}

\begin{figure}[h]
\centering
\includegraphics[width=0.35\textwidth]{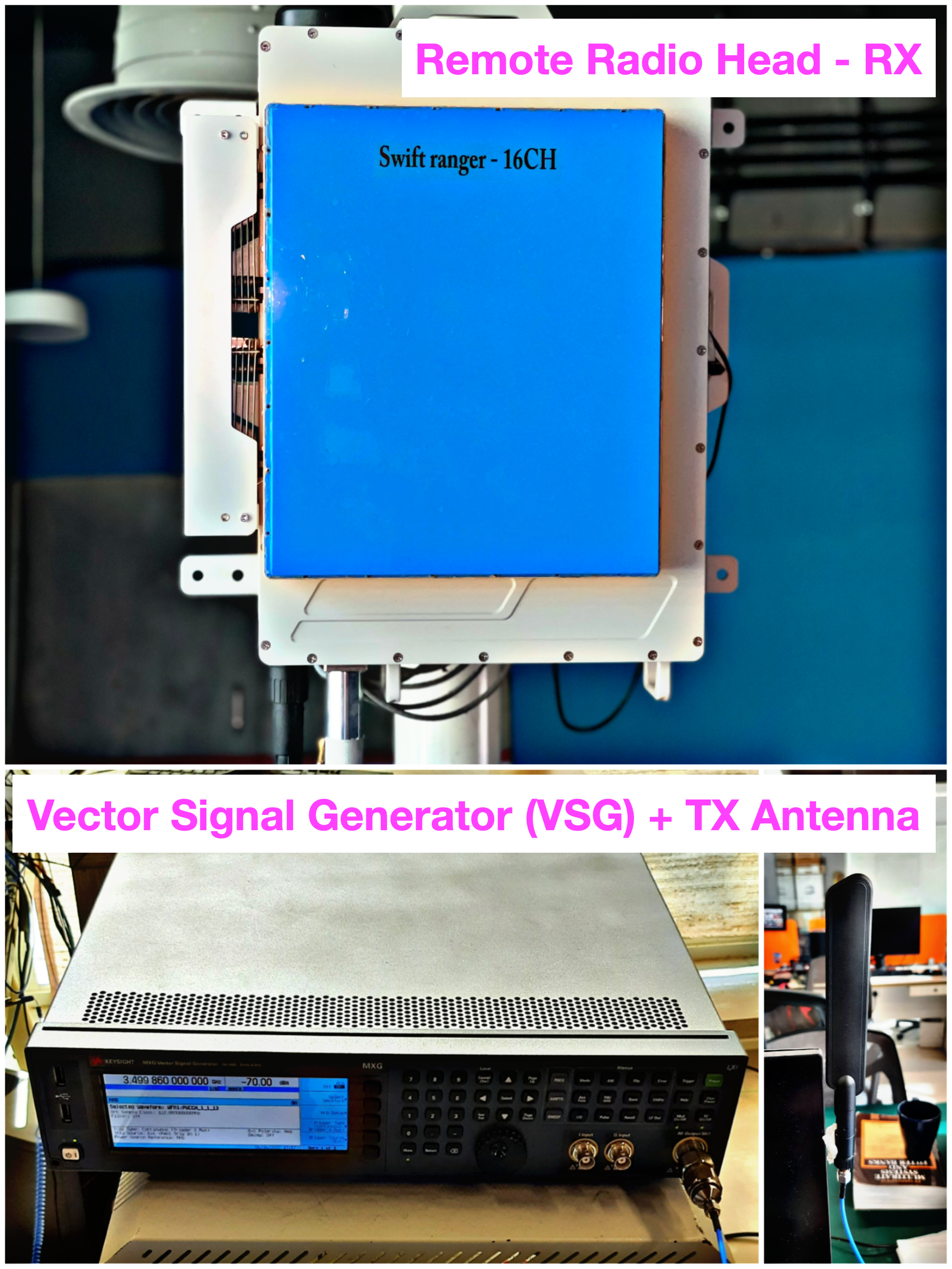}
\caption{IIT Madras 5G testbed lab setup with the Remote Radio Head used as a receiver (gNB) and the VSG used as a transmitter (Emulated UE).}
\label{fig: hw_setup}
\end{figure}

{\subsection{REAL-TIME OVER-THE-AIR DATASETS (LAB SETUP)}}
In addition to simulation datasets, we test the model using real-time over-the-air captures to assess generalizability. We use hardware captures derived from the state-of-the-art 5G testbed at IIT Madras~\cite{5gtbiitm} for more realistic testing. One of the hardware setups used (see Figure~\ref{fig: hw_setup}) consists of an N5182B Vector Signal Generator (VSG) for transmitting the  5G signal at a center frequency of $3.49986$ GHz (one of the sub-6 GHz channel raster in the n78 band). In our setup, the VSG acts like a UE, and it uses a commercial omnidirectional wideband monopole antenna to transmit the signals. The VSG is connected to the antenna through 2 SMA cables with $1.9$ dB wire loss each. 

A multi-channel 5G Remote Radio Head (RRH) with a dual-polarized antenna receives the signals over the air (for the purpose of the paper, we utilize only one antenna and one transceiver chain). The RRH operates in the n78 band with $100$ MHz bandwidth. It is ORAN-compliant and follows the 7.2b split as defined in~\cite{ORAN_spec}. Other receiver components of the RRH include an in-house Low Noise Amplifier (LNA) with 60dB gain at the receiver front end and an ADRV 9009 RF transceiver. In this indoor lab setup, we place the transmitter antenna one meter away from the receiver. The PUCCH Format 0 signal is transmitted from the VSG through its antenna, over the air, and then received at the RRH antenna, followed by the LNA and the transceiver. The signal from the 16-bit ADC is then collected and used for testing. 


{
\subsection{REAL-TIME OVER-THE-AIR DATASETS (LIVE NETWORK)}
In addition to the indoor lab setup, we also incorporate data captures obtained from an outdoor RRH  receiving PUCCH Format 0 signals from a commercial UE (Samsung A23 with a Qualcomm Snapdragon 695 modem). This RRH is part of the 5G network being run at IIT Madras in the n78 frequency band with custom developed hardware and covers the academic area of the campus.}

{The goal of this setup is to capture the PUCCH Format 0 signaling between a UE and gNB as part of a real 5G NR link, as opposed to a pre-loaded signal transmitted by the VSG in the lab setup, above. The RRH is part of an indigenous 5G NR gNB deployed on the IIT Madras campus, and is compliant to 3GPP and ORAN 7.2 split-standards. The RRH is the front-end responsible for the transmission and reception of signals. It also performs some signal processing functions like precoding, Fast Fourier Transforms (FFTs) and filtering. The other component of the gNB, the Baseband Unit (BBU), is responsible for functions like Polar Coding, Low Density Parity Check (LDPC) Coding, Channel Estimation, Channel Equalization, Cyclic Redundancy Check (CRC). Decoding of the information bits present in 5GNR control and data channels including PUCCH Format 0, is also done at the BBU. Note that, in out setup, these physical layer algorithms of the BBU are implemented on a PCIe form factor FPGA card.}

\begin{figure}[ht!]
    \captionsetup{justification=justified}
     \centering
     \begin{subfigure}[b]{0.48\textwidth}
         \centering
         \includegraphics[width=\textwidth]{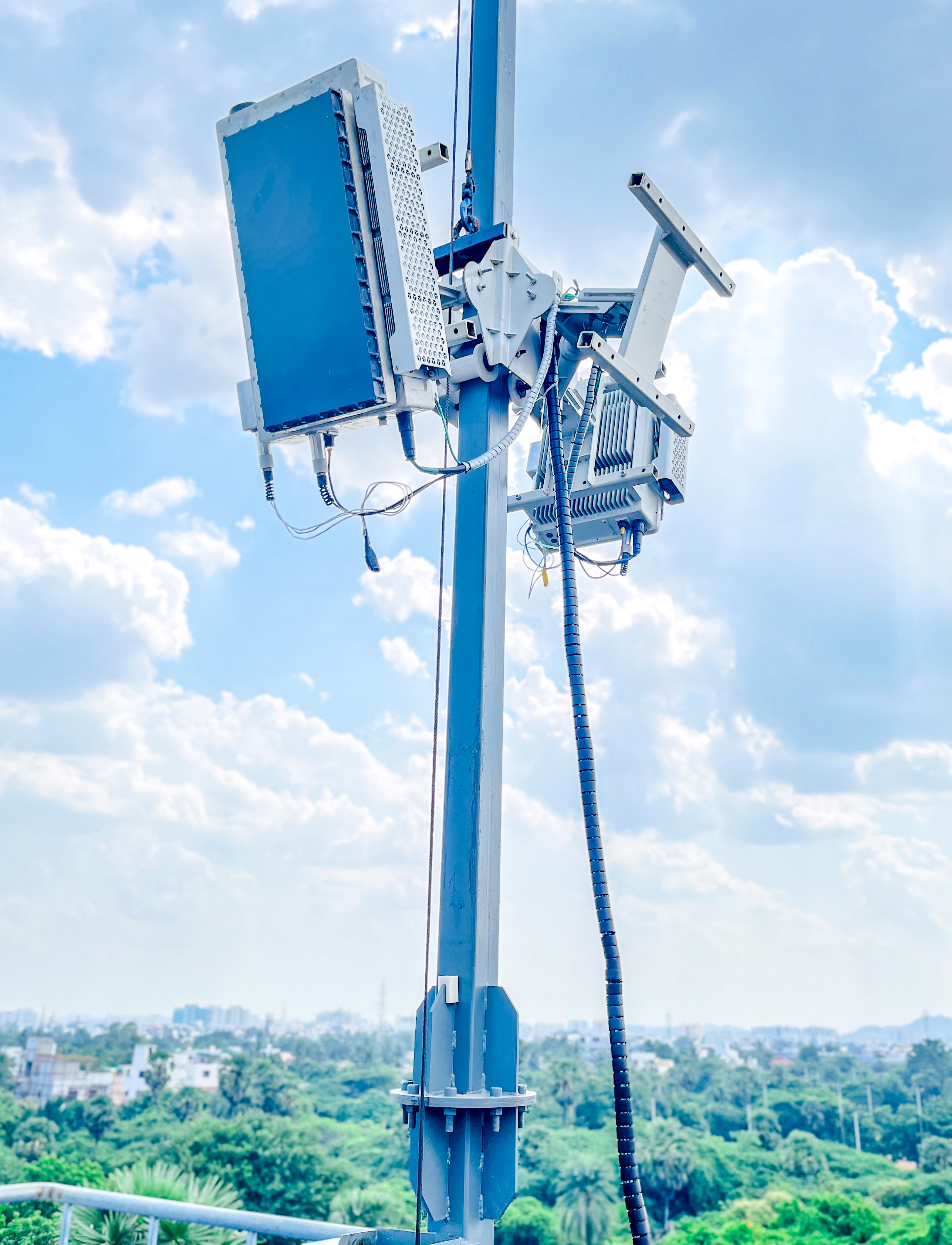}
         \caption{}
         \label{fig: RRH}
     \end{subfigure}\\
     \begin{subfigure}[b]{0.48\textwidth}
         \centering
         \includegraphics[width=\textwidth]{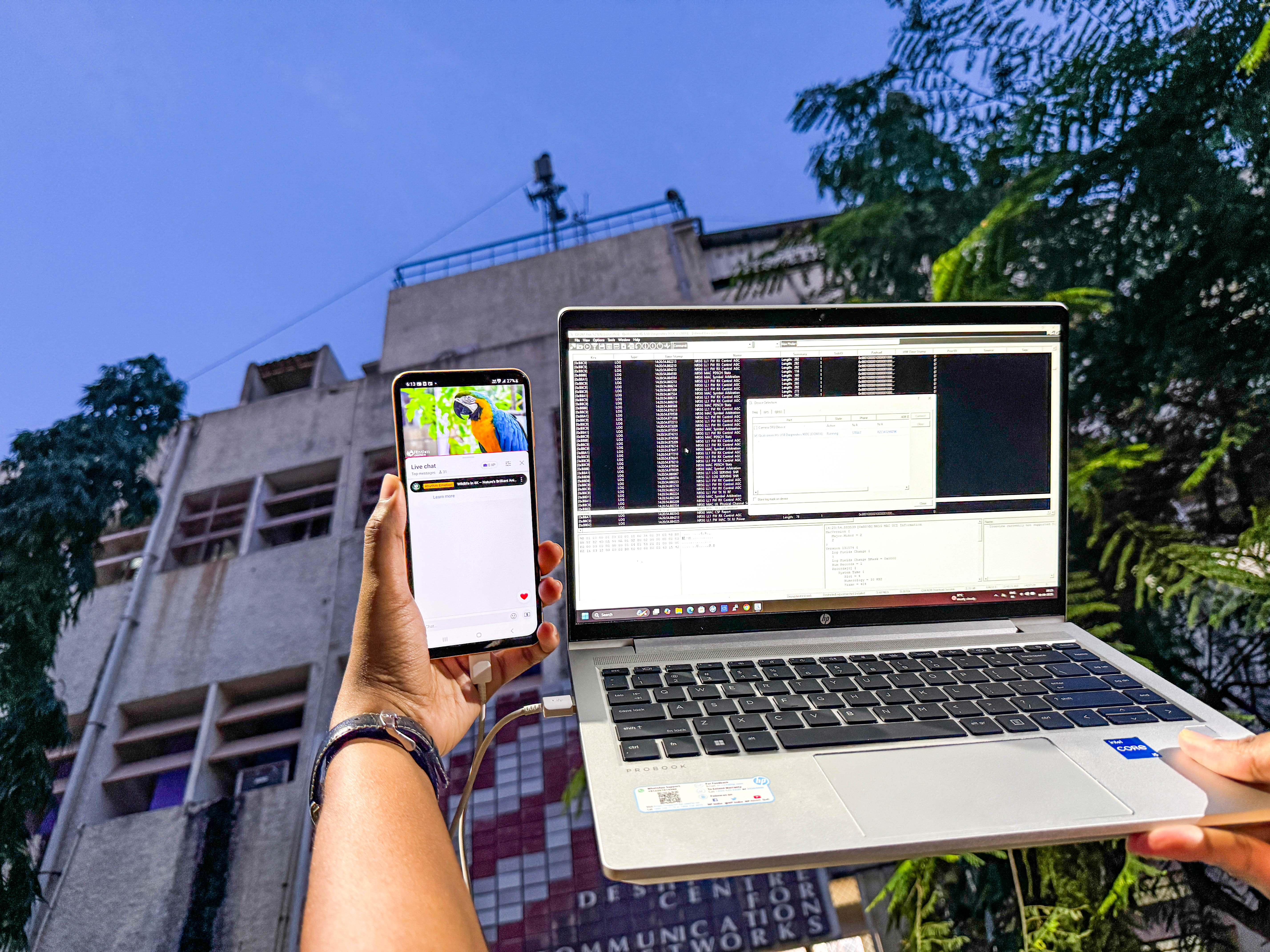}
         \caption{}
         \label{fig: UE_QXDM}
     \end{subfigure}\\
     \caption{{IIT Madras Live 5G campus Network: (a) Outdoor RRH  (16/32 channel MIMO systems) and (b) Commercial  Samsung UE connected to a laptop running QXDM.}}
     \label{fig: live_network}
\end{figure}
{To facilitate PUCCH Format 0 data capture, first, we enable data logging at both the UE and the gNB. Second, we initiate the latching procedure by which the UE decodes the Synchronization Signal Burst (SSB) and System Information Block 1 (SIB1) transmitted by the gNB. The UE then identifies itself to the gNB by transmitting a PRACH signal. This triggers the rest of the initial access procedure including both uplink and downlink control and data channel signaling, including PUCCH Format 0.}

{\subsubsection{UE SIDE DATA (GROUND TRUTH) COLLECTION}
Qualcomm eXtensible Diagnostic Monitor (QXDM) installed on a laptop connected to the UE enables collection of L1, L2 and L3 signaling logs. The saved logs are filtered to obtain only the \textit{NR5G MAC UCI Payload Information} which contains PUCCH Format 0 specific parameters such as \textit{Num HARQ bits}, \textit{Num SR bits}, \textit{HARQ Payload} and \textit{SR Payload}. These parameters can be used to generate the ground truth labels for each Format 0 instance. The logs also capture parameters such as \textit{System Frame Number (SFN)}, \textit{Slot Number}, \textit{Start Symbol}, \textit{Num Symbol} which can be used to synchronize the UCI payloads with the Format 0 signals captured at the gNB receiver.} 

{\subsubsection{gNB SIDE DATA COLLECTION}
We collect the frequency domain data received at the BBU along with the corresponding configuration parameters including \textit{SFN}, \textit{Slot Number}, number of scheduled UEs ($N_{UE}$) and \textit{Num HARQ bits}, \textit{Num SR bits} for each UE. The DFT-based correlation algorithm implemented on the BBU then decodes the UCI content from the frequency domain data. All of this information is packetized and sent to an off-the-shelf server via the PCIe interface at regular intervals. At the server, we write the data received from this PCIe interface to memory for further processing.}

{\subsubsection{DATA SYNCHRONIZATION}
It is important to synchronize the data from two independent sources - QXDM and BBU logging. This synchronization happens in two steps: 1) Coarse synchronization with the system time stamp at both ends, 2) Fine synchronization is achieved based on the \textit{SFN} and \textit{Slot Number} embedded in the packets.  Once synchronized, we can map the ground truth labels (UCI transmitted by the UE) to the appropriate PUCCH Format 0 signal received by the gNB and the UCI content decoded by DFT-based conventional algorithm (for comparison).}

\section{MODEL TRAINING AND TESTING}
\label{model_training_and_testing}
In this section, we describe the model training and testing process including forward and backward propagation, optimizer and loss, as well as the specific datasets used for training and testing. The training and testing were done using Python and TensorFlow on an NVIDIA A100 Tensor Core GPU. The code is available at~\cite{aiml_pucch_dataset} for download.
\subsection{MODEL TRAINING}
{During training, we employ dropout~\cite{srivastava2014dropout} with a probability of 0.5, for all the models described in this paper. Dropout achieves both ensemble learning~\cite{hara2016analysis} and regularization~\cite{Goodfellow-et-al-2016} by ``dropping out" or cutting off certain neurons and their connections randomly with a certain probability. During each forward pass, the sigmoid activation at the output layer generates probabilities which are used to compute a binary cross entropy loss. The loss is then back-propagated using Stochastic Gradient Descent (SGD) with a learning rate of $10^{-2}$ and a momentum parameter of $0.9$.}

Based on the insights gained from our previous work~\cite{yerrapragada2023machine}, we train the neural network using data with a middle SNR of $10$ dB. We have found that this model achieves good generalization across the entire range of SNR during testing, i.e., $10$ dB SNR is sufficient enough for the model to learn the true patterns both in the underlying data as well as noise. Training data also includes a $\Delta$ value of $2$ (middle value of metadata offset), all UE multiplexing scenarios ($\Tilde{N}_{UE} = 0, 1, \dots 12$), the TDLC300 channel model, and Doppler frequency shifts of $0, 500, 1000, 1500$, and $2000$ Hz. $75\%$ of the data is used for training, from which a further 30\% is used for validation and fine-tuning of hyperparameters. We train the model for 150 epochs (an epoch is 1 pass of the entire dataset through the neural network). The training dataset size is $1638K$.

{\subsection{MODEL TESTING}
\label{model_testing}
To ensure consistent performance of the model, we test it across a wide range of parameters using both simulated and hardware datasets. For the simulated data, the testing instances span SNRs of $0, 2, 4, \dots 20$dB, TDLC300, TDLC30, TDLA300, TDLA30 channel models, Doppler frequency shifts of $0,400,800,1200,1600$ and $2000$ Hz, all UE multiplexing scenarios ($\Tilde{N}_{UE} = 0, 1, \dots 12$) and $\Delta$ values of $0$, $2$ and $4$. For the hardware data, the testing instances span SNRs $0,5,10,15,$ and $20$ dB,  all UE multiplexing scenarios ($\Tilde{N}_{UE} = 0, 1, \dots 12$) and and $\Delta$ values of $0$, $2$ and $4$. For the live network data, a single UE connected to the QXDM tool is subjected to different SNRs ($1$, $3$, $7$, and $10$ dB), and the data is collected at the gNB receiver.}

{Based on the known scheduling information from L2, the superset of all possible cyclic shifts a UE can transmit can be pre-determined, as shown by the $\overline{\alpha}$ examples in Section~\ref{subsec: mathematical_model_pucch_f0}. Even in the case of a multiplexed UE scenario, the possible outputs of the neural network (\textit{alpha indicators}) can be derived by combining all the individual UE's possible cyclic shifts. This information, in the form of a binary mask vector is applied to the outputs of the models, during inference. The function of the mask is to nullify the specific cyclic shift predictions that are apriori known to be incorrect. We also observe a significant gain when the inference mask is applied to the neural network's prediction.}

{For example, when $2$ UEs with each of them being scheduled with a $1$ SR and $1$ HARQ transmission, a total of $4$ cyclic shifts could be transmitted individually. So, the binary mask will assert all the $8$ possible cyclic shifts as the inference mask to the output of the neural network.}

\section{RESULTS AND DISCUSSION}\label{sec: results}
In this section, we describe the performance results of the UCINet0 NN architecture shown in Figure~\ref{fig: nn_arch}. 

\subsection{PERFORMANCE METRICS AND THEIR MOTIVATION}
We use model test accuracy as the main performance metric. We also show confusion matrices to indicate the distribution of model predictions. Wherever appropriate, we show comparisons of the model performance with the corresponding performance shown by the DFT-based correlation algorithm. 

Accuracy is simply the number of correct predictions in relation to total predictions. Though it is a useful high-level metric to gauge model performance, in cases where the classes in a dataset are unequally distributed or when, as is the case in this paper, there are more than 2 classes, accuracy alone as a metric for an NN classifier does not offer the full picture of model performance. It is helpful to know if the model classifies all classes equally well or if it is more ``confused" by certain classes compared to others. Hence, calculating a confusion matrix gives a better insight into any patterns that may exist in both correct and wrong classifications made by the model. In this paper, we show 3 types of confusion matrices: (1) A multi-label confusion matrix that shows how well each $\alpha$ is predicted; (2) A confusion matrix showing how well the model predicts the correct number of multiplexed UEs, and (3) A confusion matrix showing how each value of $\alpha$ (including the cases when no $\alpha$ is selected) is classified. We present this last metric in the form of a column chart for ease of visualization. These 3 types of confusion matrices provide a sense of interpretability of the inner workings of the NN model. 

\subsection{ACCURACY AND LOSS} \label{subsec: accuracy}
\begin{figure}[ht!]
    \captionsetup{justification=justified}
     \centering
     \begin{subfigure}[b]{0.35\textwidth}
         \centering
         \includegraphics[width=\textwidth]{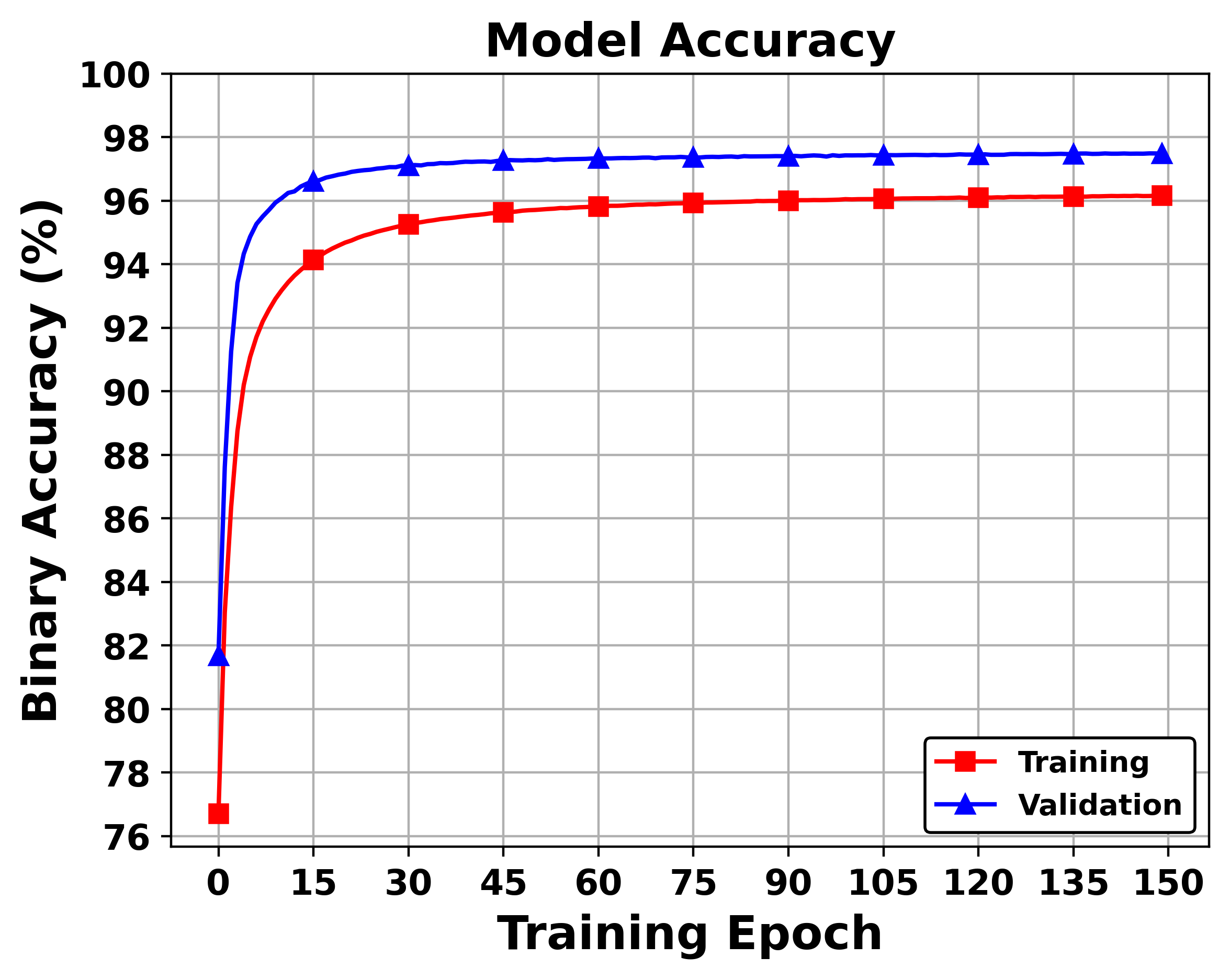}
         \caption{}
         \label{fig: acc_vs_epoch}
     \end{subfigure}
     \\
     \begin{subfigure}[b]{0.35\textwidth}
         \centering
         \includegraphics[width=\textwidth]{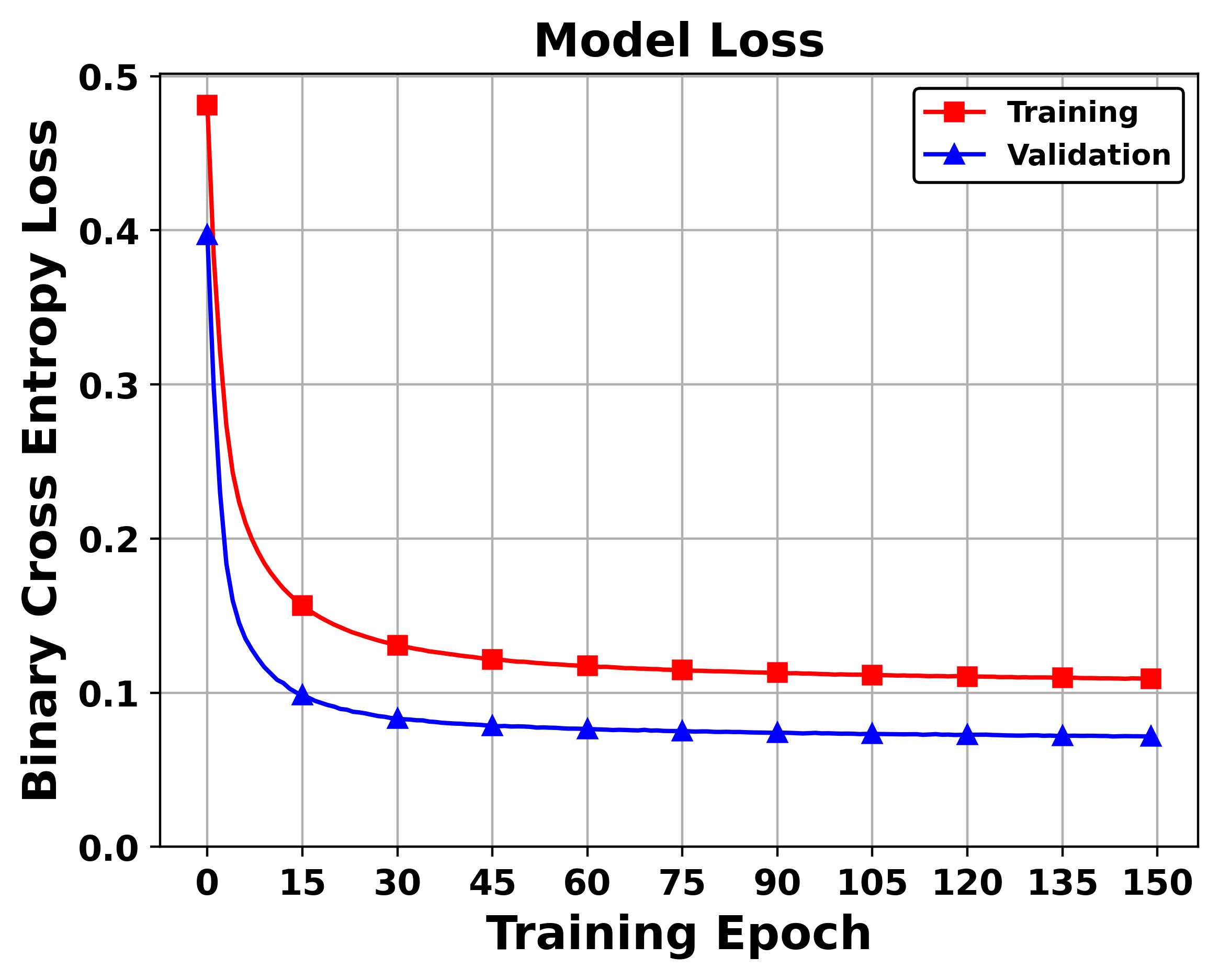}
         \caption{}
         \label{fig: loss_vs_epoch}
     \end{subfigure}
        \caption{{Training was performed only using simulated data at SNR = $10$ dB. (a) Model Accuracy vs. Epoch for simulated data. (b) Model Loss vs. Epoch for simulated data.}}
        \label{fig: acc_loss_vs_epoch}
\end{figure}

Figure~\ref{fig: acc_vs_epoch} shows the training and validation accuracy with respect to the training epochs. Figure~\ref{fig: loss_vs_epoch} shows the training and validation loss. Both curves follow the expected pattern for a well-fit model. The accuracy gradually increases and the loss steadily decreases with epoch and eventually converges. We also note that the validation accuracy and loss values are slightly better than the corresponding training values. One possible explanation for this could be the use of dropout. As stated above, a dropout with probability 0.5 drops neurons $50\%$ of the time during training. However, during validation and testing, the entire model is used, leading to a higher accuracy/lower loss. 

\begin{figure}[ht!]
    \captionsetup{justification=justified}
     \centering
     \begin{subfigure}[b]{0.48\textwidth}
         \centering
         \includegraphics[width=\textwidth]{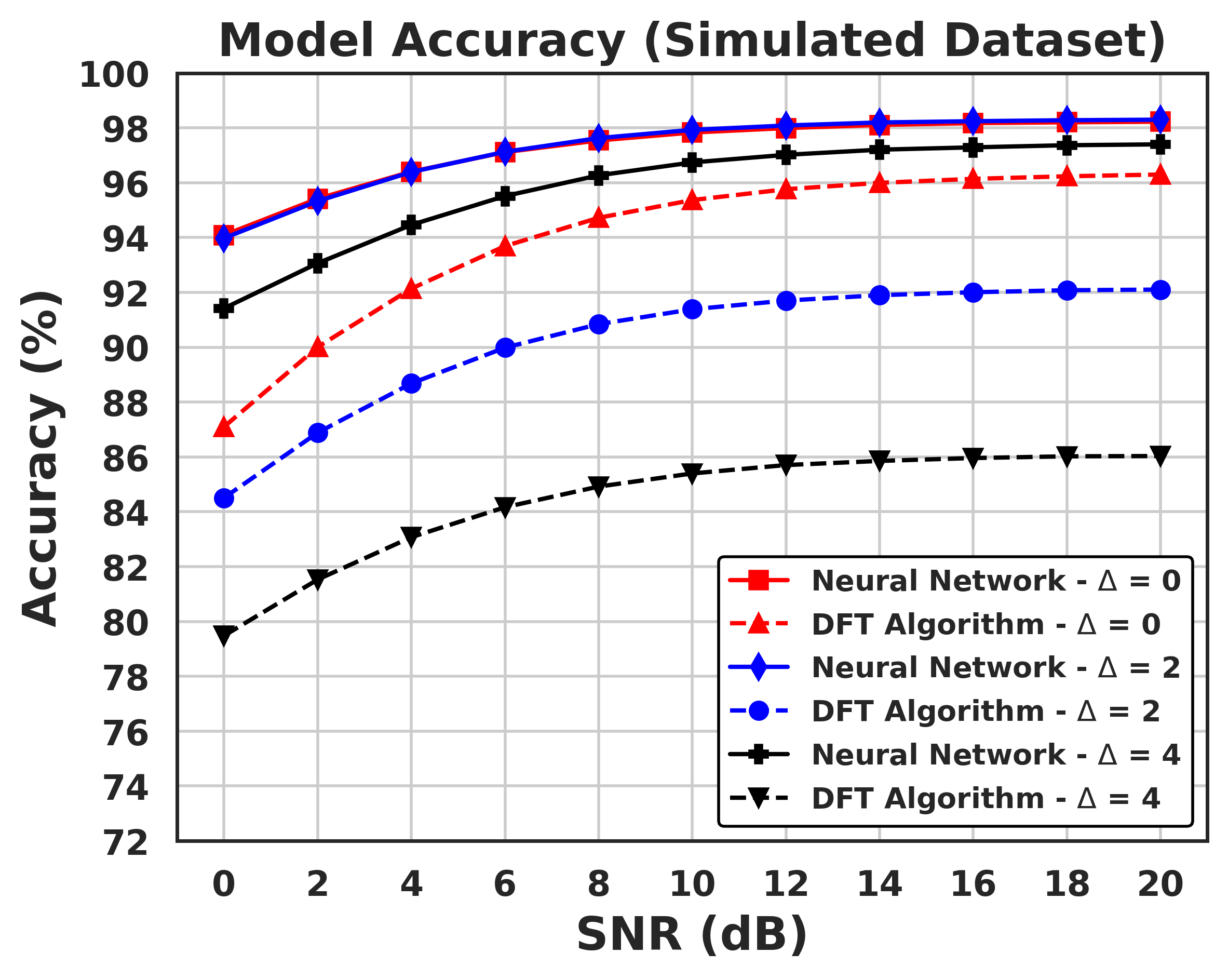}
         \caption{}
         \label{fig: acc_vs_snr_sim}
     \end{subfigure}
     \\
     \begin{subfigure}[b]{0.48\textwidth}
         \centering
         \includegraphics[width=\textwidth]{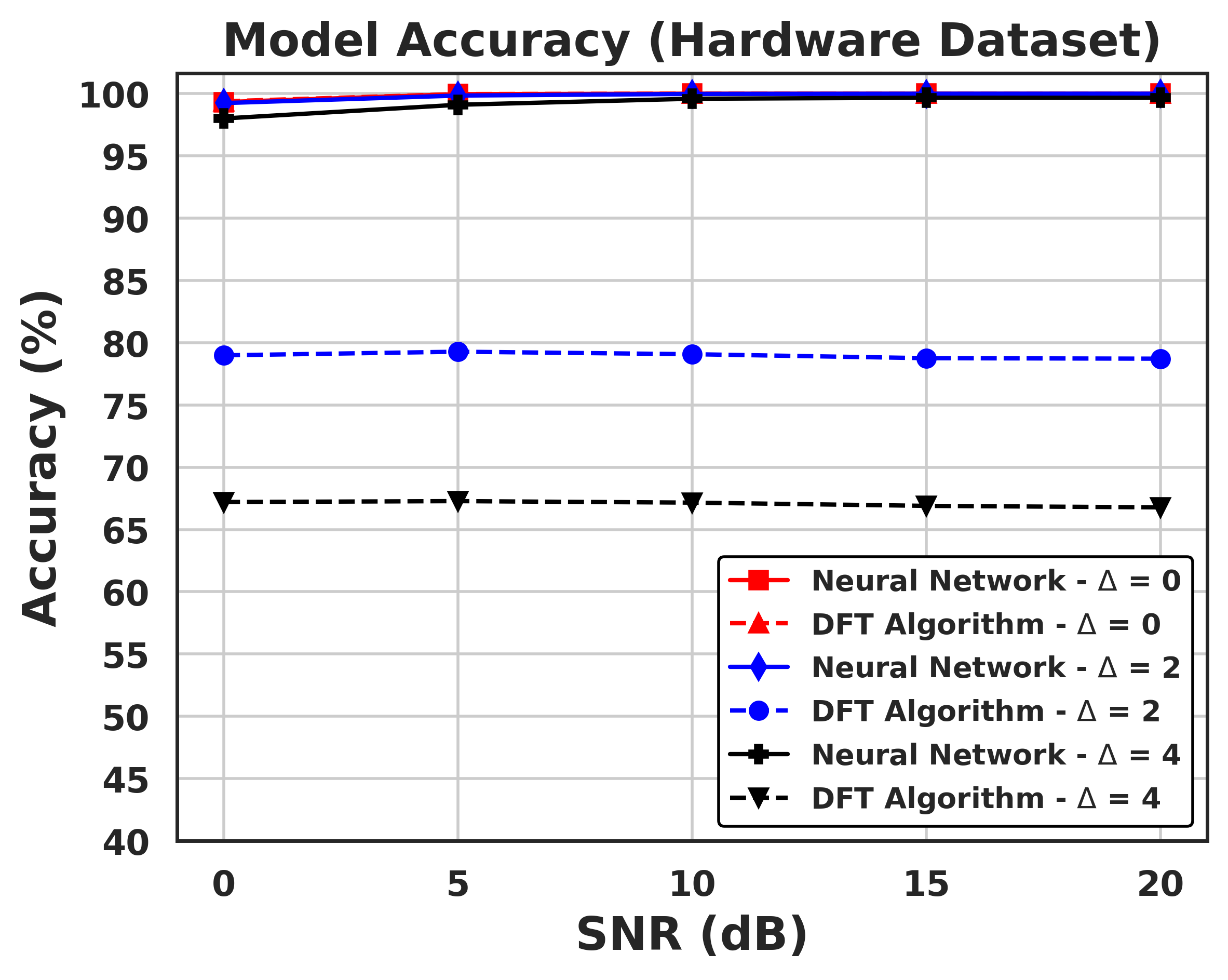}
         \caption{}
         \label{fig: acc_vs_snr_hw}
     \end{subfigure}
        \caption{{Testing Accuracy vs SNR for (a) simulated data and (b) hardware-captured data (lab setup). The test dataset for each SNR contains a combination of $\Tilde{N}_{UE} = \{0, 1, 2, \dots, 12\}$}}
        \label{fig: acc_vs_snr}
\end{figure}

Figure~\ref{fig: acc_vs_snr_sim} shows the test accuracy of the model versus SNR for simulated data. Figure~\ref{fig: acc_vs_snr_hw} shows the test accuracy of the model versus SNR for hardware-captured data. In both cases, the test dataset for each SNR contains a combination of all possible values of $\Tilde{N}_{UE} = \{0, 1, 2, \dots, 12\}$. Both Figures~\ref{fig: acc_vs_snr_sim} and~\ref{fig: acc_vs_snr_hw} show that model accuracy increases with SNR. For the simulated data, the model significantly outperforms the DFT-based correlation algorithm for all values of $\Delta$.

For the hardware-captured data, when $\Delta = 0$, the model performs slightly better than the DFT-based correlation algorithm. However, as the SNR increases, the accuracy of the model and that of the DFT-based correlation algorithm converge. This is because $\Delta = 0$ constitutes a best-case scenario in which both algorithms are aided by the fact that the receiver knows exactly how many UEs are multiplexed. At higher values of $\Delta$, the convergence is not seen, and the NN model shows a significant gain compared to the DFT-based correlation algorithm. 

Furthermore, from Figures~\ref{fig: acc_vs_snr_sim} and~\ref{fig: acc_vs_snr_hw}, we observe an interesting difference between the performance of the neural network with simulated data versus hardware-captured data. For all values of $\Delta$, there is a significant SNR gain with the hardware-captured data compared to simulated data. For example, when $\Delta = 0$, the accuracy with hardware-captured data reaches 100\% at as low an SNR as $5$ dB. In contrast, the accuracy with simulated data requires more than $15$dB SNR to reach 98\%). This difference could be attributed to the characteristics of the datasets themselves. The simulated dataset includes a strong fading component (both slow and fast) that impairs the transmitted signal. However, owing to the short propagation distance and stationary VSG, such a strong fading component is absent in the current 5G testbed setup that was used to obtain the hardware captures. We note that the captures still incorporate other real-world impairments caused by front-end hardware elements like filters, DACs, ADCs, up/down converters, and amplifiers. In summary, the hardware dataset models impairments due to front-end elements well. It also models over-the-air impairments but does not encapsulate all characteristics of a typical communication link between a gNB and UE that might span hundreds of meters. On the other hand, the simulated data models over-the-air characteristics well, but it doesn't model impairments due to hardware elements. We also point out that the IIT Madras 5G Testbed is moving towards a full-fledged deployment of Remote Radio Heads at on-campus sites, which will allow the inclusion of even more realistic field data into future work.

Another observation from Figures~\ref{fig: acc_vs_snr_sim} and~\ref{fig: acc_vs_snr_hw} is that the DFT-based correlation algorithm performs worse on hardware data, showing that it is negatively affected not only by impairments from hardware elements but also by even the slightest over-the-air fading. On the other hand, the NN Model outperforms the DFT method in all scenarios of hardware and simulated datasets.

\begin{figure}[ht!]
    \captionsetup{justification=justified}
     \centering
     \includegraphics[width=0.48\textwidth]{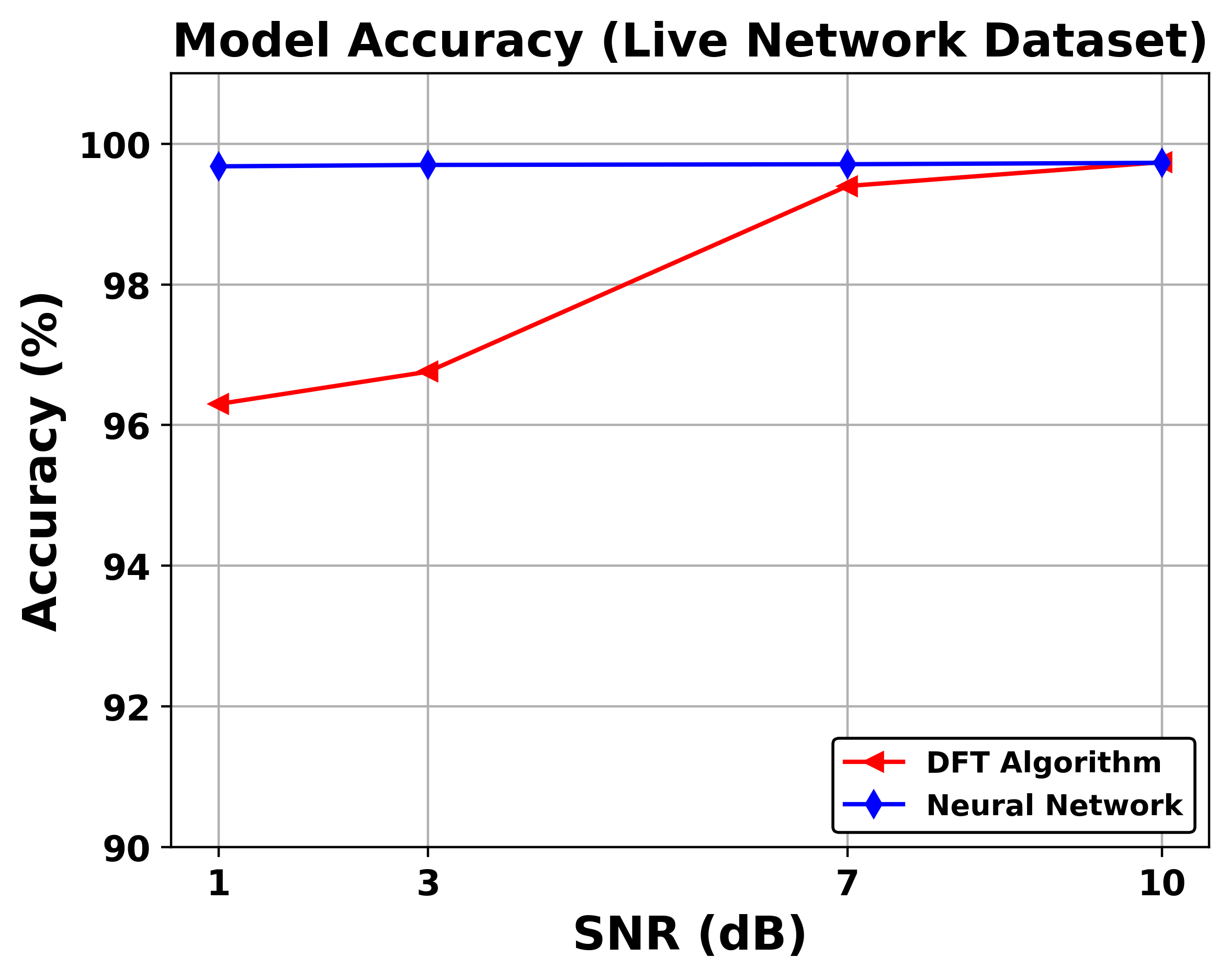}
     \caption{{Testing Accuracy vs average SNR (approximate) for data captured from the live network. These average SNRs have been obtained at different locations of the network. }}
     \label{fig: acc_vs_snr_live_network}
 \end{figure}

Figure~\ref{fig: acc_vs_snr_live_network} exhibits the model performance for the data captured from a live network. We observe that the model's performance remains consistent across SNRs, reflecting the robustness of the model in live network signalling.

\begin{figure}[ht!]
    \captionsetup{justification=justified}
     \centering
     \begin{subfigure}[b]{0.48\textwidth}
         \centering
         \includegraphics[width=\textwidth]{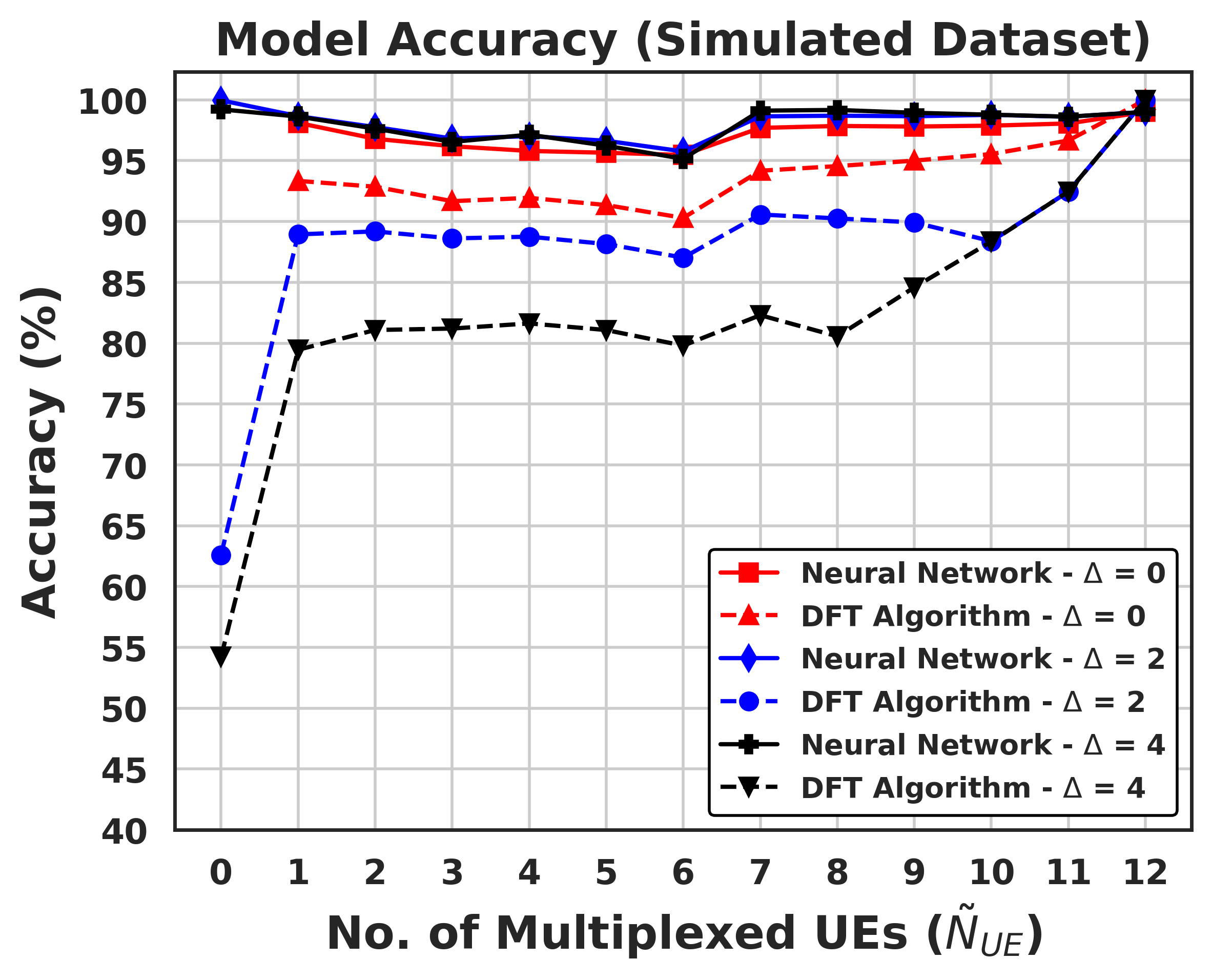}
         \caption{}
         \label{fig: acc_vs_mux_ue_sim}
     \end{subfigure}
     \\
     \begin{subfigure}[b]{0.48\textwidth}
         \centering
         \includegraphics[width=\textwidth]{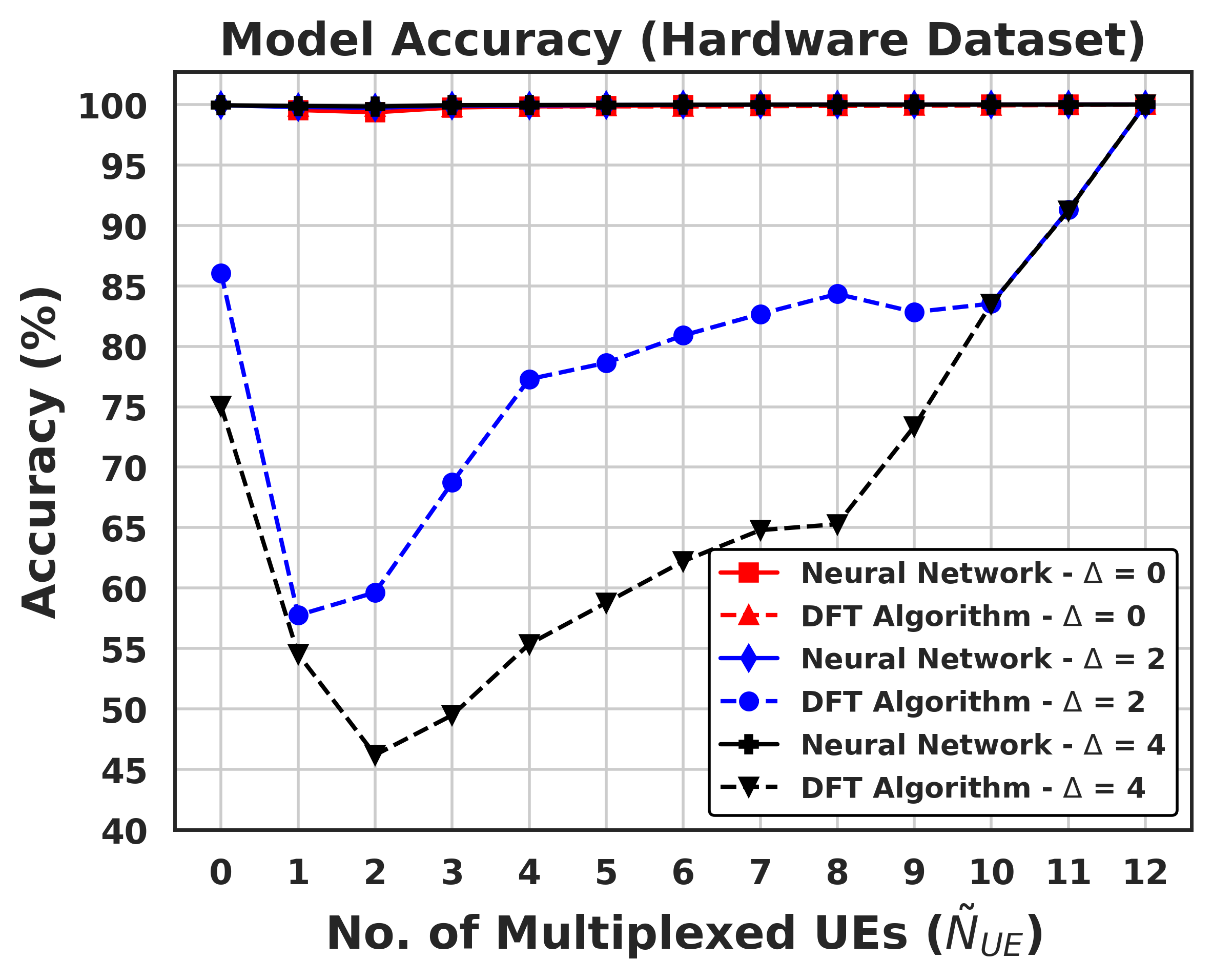}
         \caption{}
         \label{fig: acc_vs_mux_ue_hw}
     \end{subfigure}
        \caption{{Testing Accuracy vs Number of Multiplexed UEs ($\Tilde{N}_{UE}$) for (a) simulated data and (b) hardware-captured data. The test dataset for each value of $\Tilde{N}_{UE}$ contains a combination of SNR = $\{0, 2, 4, \dots, 20\}$ dB for (a) and SNR = $\{0, 5, 10, 15, 20\}$ dB for (b).}}
        \label{fig: acc_vs_mux_ue}
\end{figure}

Figure~\ref{fig: acc_vs_mux_ue_sim} shows the test accuracy of the NN model versus the number of multiplexed UEs ($\Tilde{N}_{UE}$) for simulated data. Figure~\ref{fig: acc_vs_mux_ue_hw} shows the test accuracy of the NN model versus the number of multiplexed UEs ($\Tilde{N}_{UE}$) for hardware-captured data. In both cases, the test dataset for each $\Tilde{N}_{UE}$ contains a combination of all possible values of SNR $= \{0, 5, 10, 15, 20\}$ dB. It is clear that the accuracy decreases as the maximum offset $\Delta$ increases. Note that no value is plotted when $\Tilde{N}_{UE} = 0$ and $\Delta=0$ since this case corresponds to the scenario in which no UEs are transmitting on a given Resource Block and the receiver knows apriori that no UEs are scheduled as well. In this case, the receiver need not even run. When $\Tilde{N}_{UE} = 12$, all the multiplexed UEs transmit only SRs. For both the simulated and hardware datasets, the DFT-based correlation algorithm selects all 12 values from the 12-point DFT and assigns one value to each UE. Since all UEs are transmitting SRs, the order of assignment doesn't matter, and hence the accuracy is always 100\%. The NN model also seems to recognize that the only scenario in which 12 UEs can be multiplexed is when they all transmit exactly the same information, i.e., SRs. 


\begin{figure}[ht!]
    \captionsetup{justification=justified}
     \centering
     \includegraphics[width=0.48\textwidth]{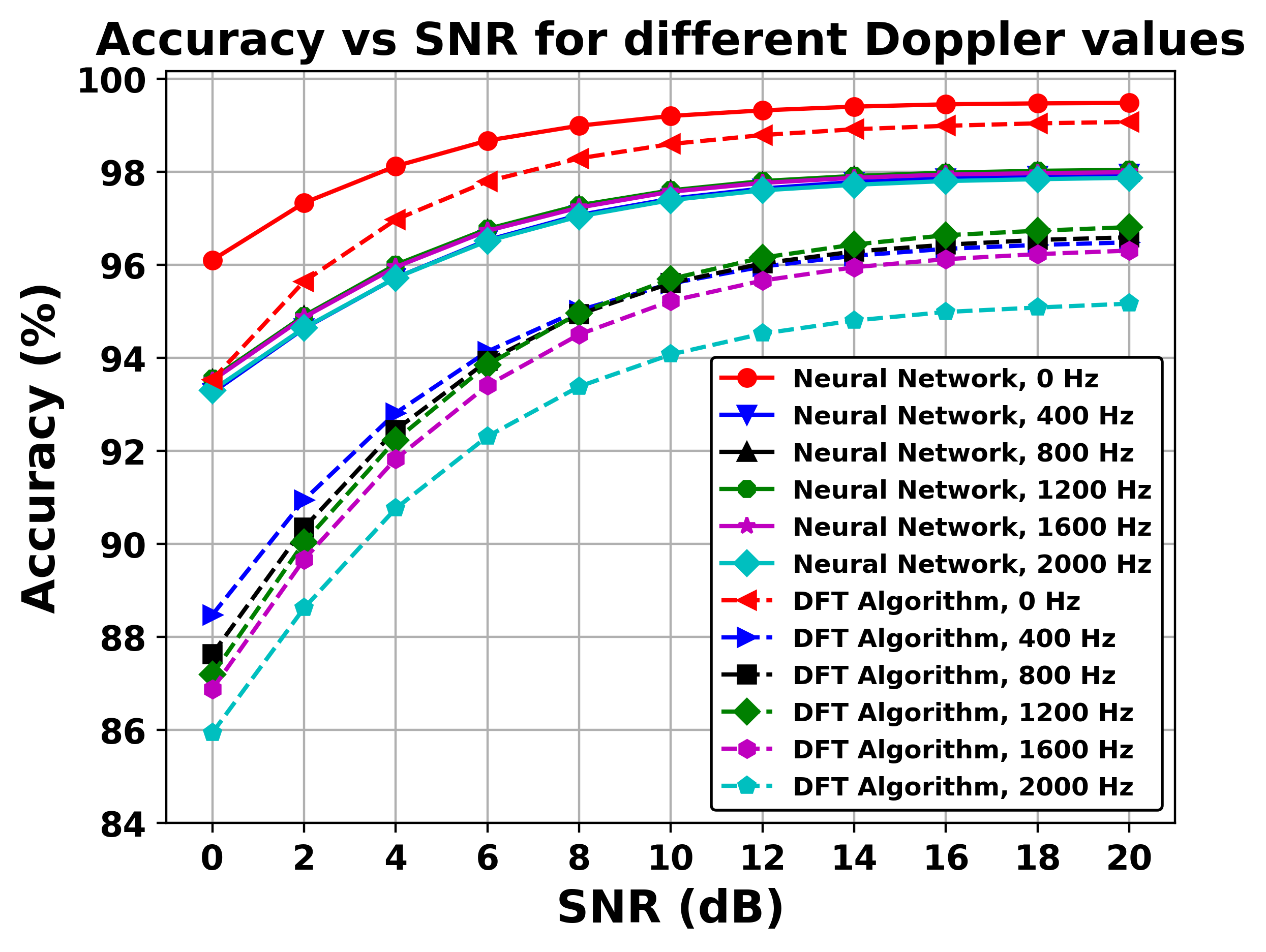}
     \caption{{{Accuracy versus SNR graph for multiple Doppler frequency shifts ranging from 0 to 2000 Hz.}}}
     \label{fig: acc_vs_snr_fd_plot}
 \end{figure}

Figure~\ref{fig: acc_vs_snr_fd_plot} shows the test accuracy of the NN model versus SNR, for simulated data, across multiple Doppler frequency shifts. The graph compares both the NN and the conventional DFT-based correlation algorithm. We observe that across SNRs, for any Doppler frequency shift, the neural network based decoder performs consistently better than the conventional approach. The variation in accuracy values for different Doppler frequency shifts is minimal for the NN based decoder when compared to the conventional decoder.

\begin{figure}[ht!]
    \centering
    \includegraphics[width=0.48\textwidth]{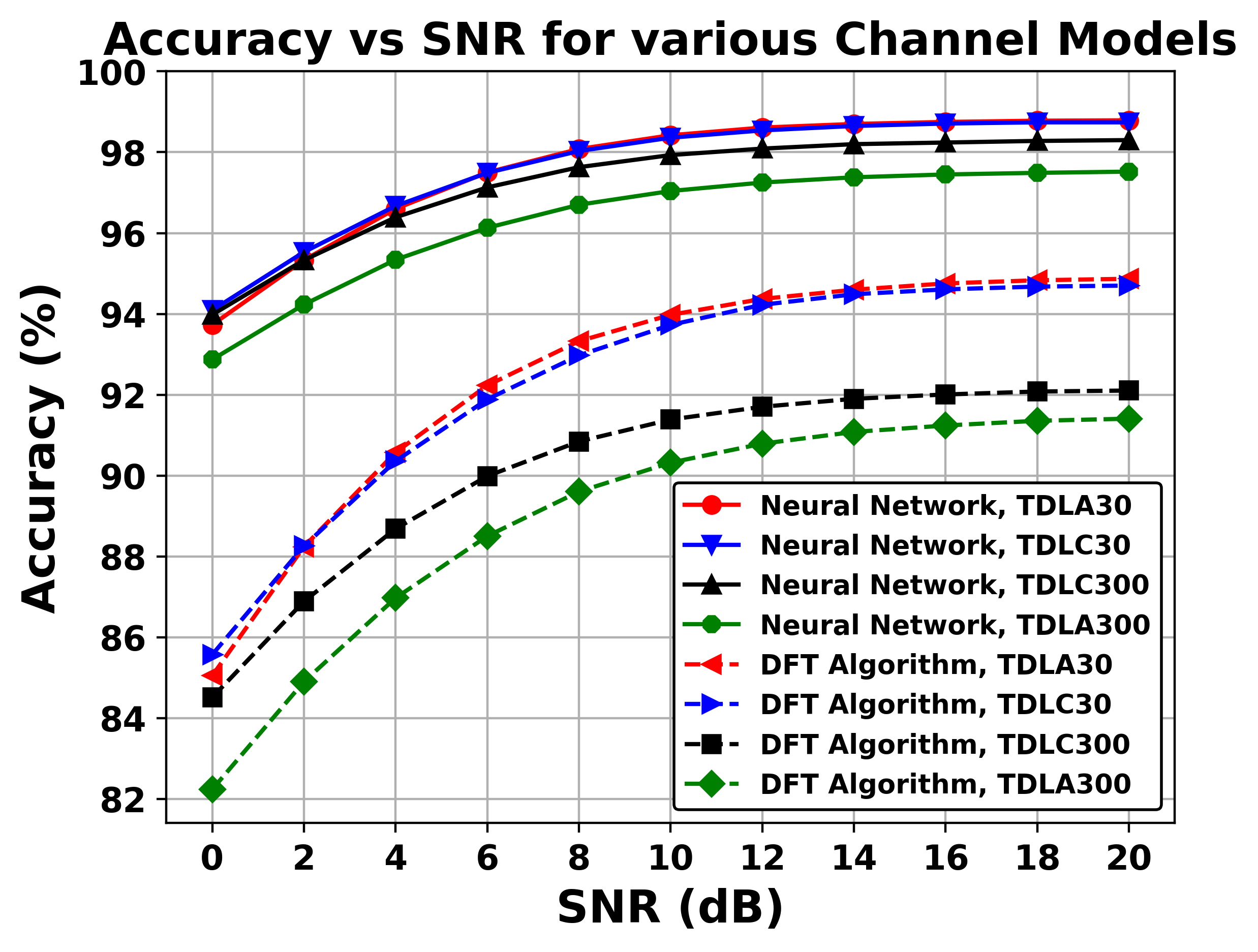}
    \caption{{{Accuracy vs SNR for multiple fading channel scenarios}}}
    \label{fig: acc_vs_snr_channel}
\end{figure}

{Figure~\ref{fig: acc_vs_snr_channel} shows the test accuracy of the NN model versus SNR, for simulated data, across multiple channel models. The graph compares both the NN and the conventional DFT-based correlation algorithm. Noting that the model has been trained exclusively on the TDLC300 channel model, we observe that for delay spreads lower than 300 ns, for both TDLA and TDLC, the performance is equivalent to that of TDLC300. For TDLA300, there is a performance drop of approximately $1\%$ across all SNRs. While this can be considered to be an acceptable generalization performance across various channel models, the $1\%$ gap can easily be reduced by including some instances of TDLA300 in the training.} 

\subsection{CONFUSION MATRICES AND COLUMN CHARTS}
In this subsection, confusion matrices and column charts are presented to gain deeper understanding of the model's performance.
Considering the relatively low variance in accuracy across various Doppler frequency shifts, as shown in Figure~\ref{fig: acc_vs_snr_fd_plot}, further analysis incorporates all the Doppler frequency shifts together in the same dataset.

\begin{figure}[ht!]
    \captionsetup{justification=justified}
     \centering
     \includegraphics[width=0.48\textwidth]{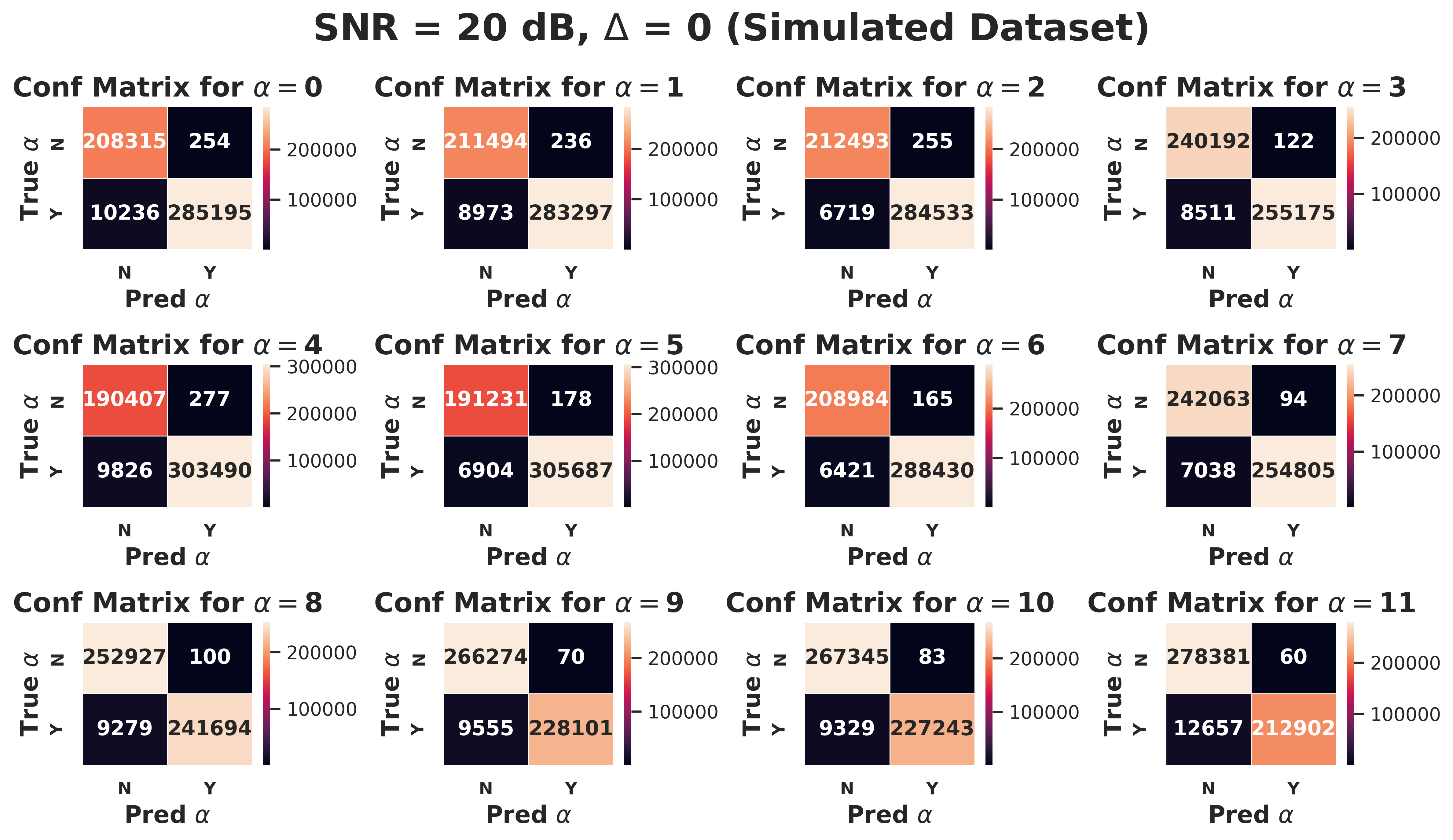}
    \caption{{Multi-label confusion matrix at SNR = $20$dB and $\Delta$ = $0$ for simulated test dataset}}
    \label{fig: cm_multilabel_snr_20_off_0}
\end{figure}

\begin{figure}[ht!]
    \captionsetup{justification=justified}
     \centering
     \includegraphics[width=0.48\textwidth]{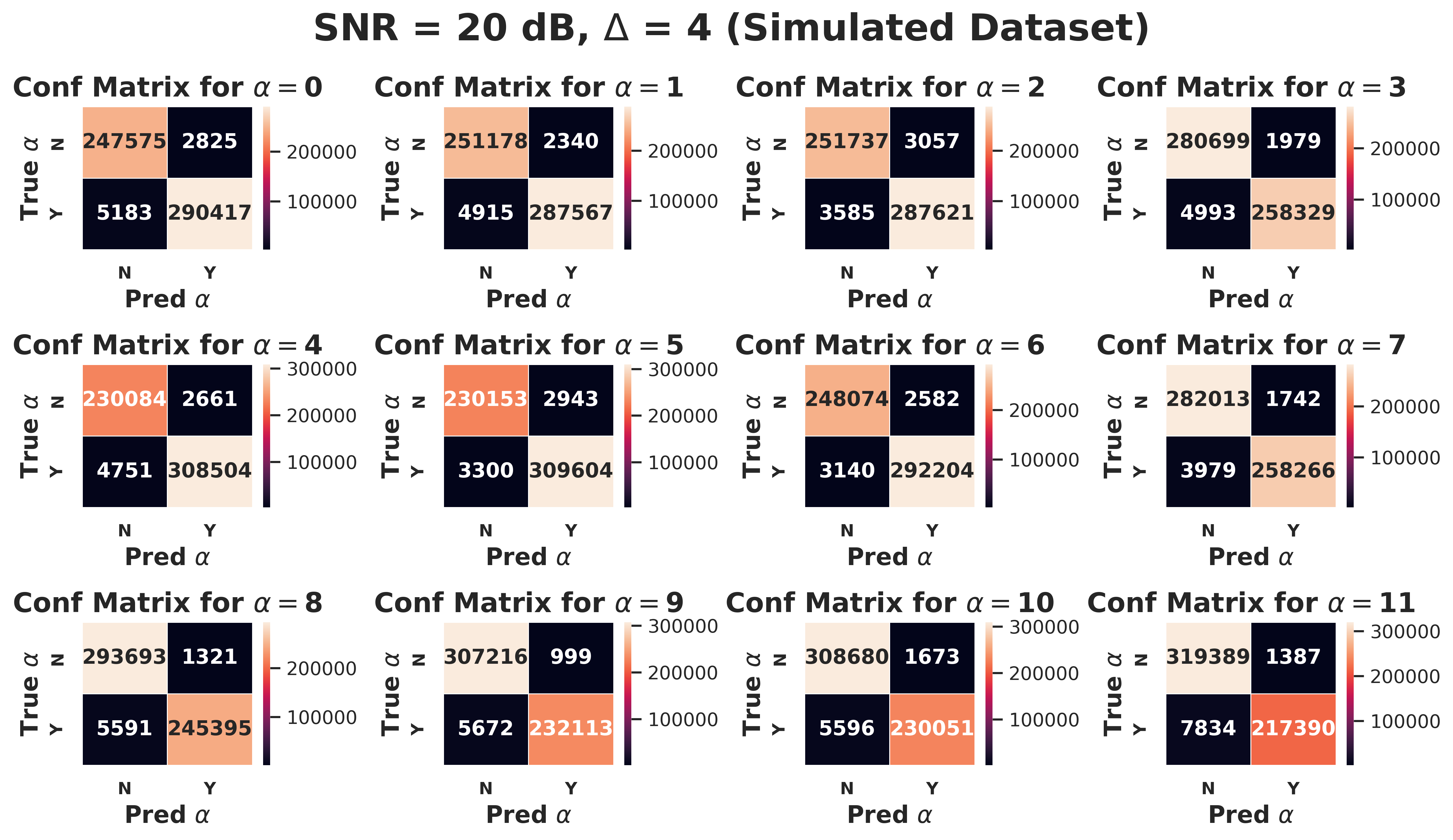}
    \caption{{Multi-label confusion matrix at SNR = $20$dB and $\Delta$ = $4$ for simulated test dataset}}
    \label{fig: cm_multilabel_snr_20_off_4}
\end{figure}

The confusion matrices in Figures~\ref{fig: cm_multilabel_snr_20_off_0} and~\ref{fig: cm_multilabel_snr_20_off_4} present the overall performance of the multi-label classifier model. Figures~\ref{fig: cm_multilabel_snr_20_off_0} and~\ref{fig: cm_multilabel_snr_20_off_4} show the multi-label confusion matrix for the best ($\Delta = 0$) and worst ($\Delta = 4$) cases of $20$dB SNR, respectively, for simulated data. Multi-label classification can be thought of as a combination of several independent binary classifiers. Diagonally dominant confusion matrices indicate that each of the binary classifiers is well-trained. For a given SNR, as $\Delta$ increases, the model leans more towards false positives because the model (whose 25th input is $N_{UE} \in [\Tilde{N}_{UE}, \Tilde{N}_{UE} + \Delta]$) is trying to predict $\Tilde{N}_{UE}$ number of UEs. The value of $\Delta$ is always positive, meaning that the number of expected UEs, as indicated by L2, is always greater than or equal to the actual number of UEs transmitting on a given resource allocation. Hence, the model almost always over-predicts rather than missing a user's detection. 

\begin{figure}[ht!]
    \captionsetup{justification=justified}

     \centering
     \includegraphics[width=0.48\textwidth]{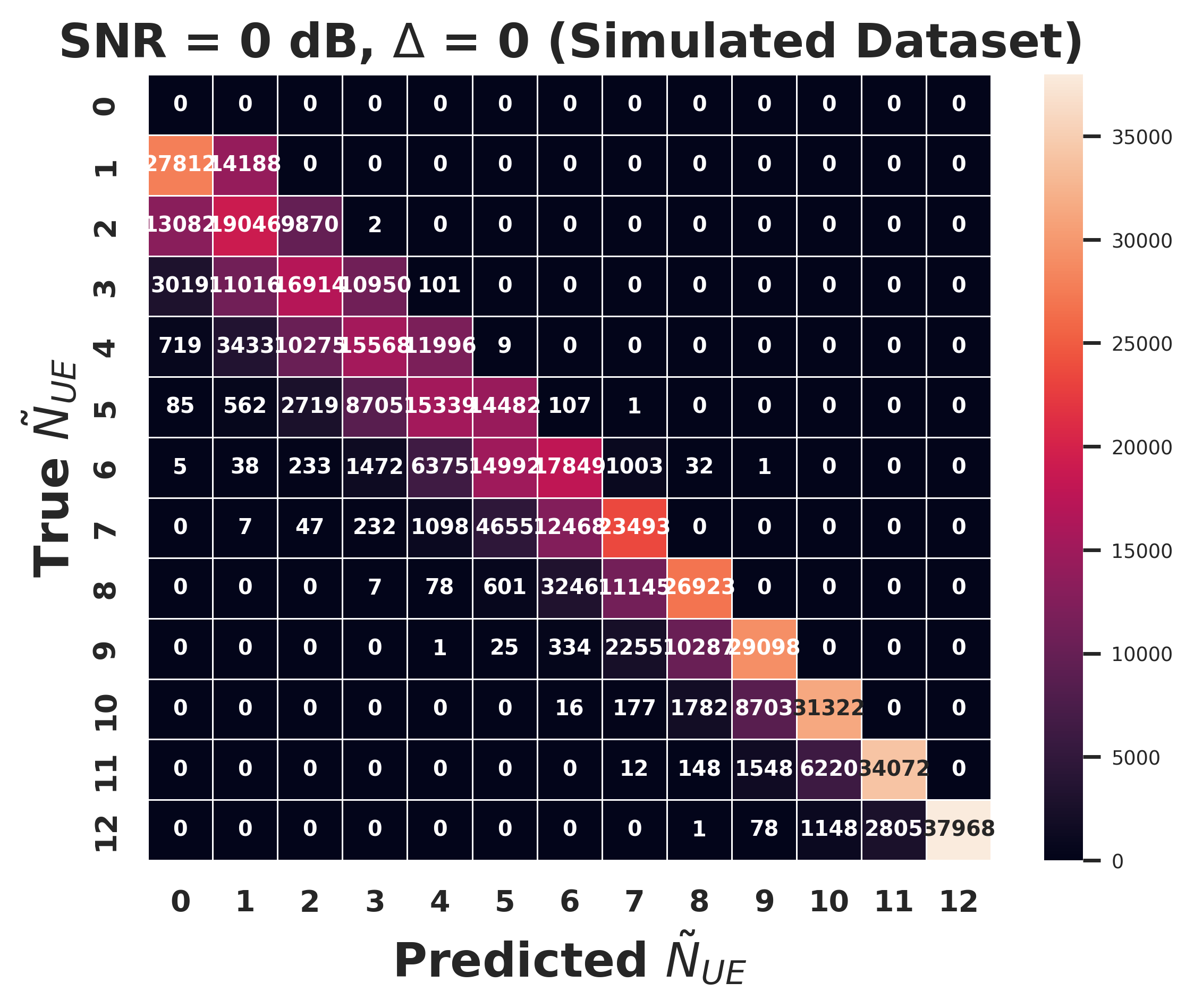}

    \caption{{Confusion matrix for the number of multiplexed UEs $\Tilde{N}_{UE}$, at SNR = $0$ dB and $\Delta$ = $0$ for simulated test dataset.}}
    \label{fig: cm_num_UE_snr_0_off_0}
\end{figure}

\begin{figure}[ht!]
    \captionsetup{justification=justified}
     \centering

     \includegraphics[width=0.48\textwidth]{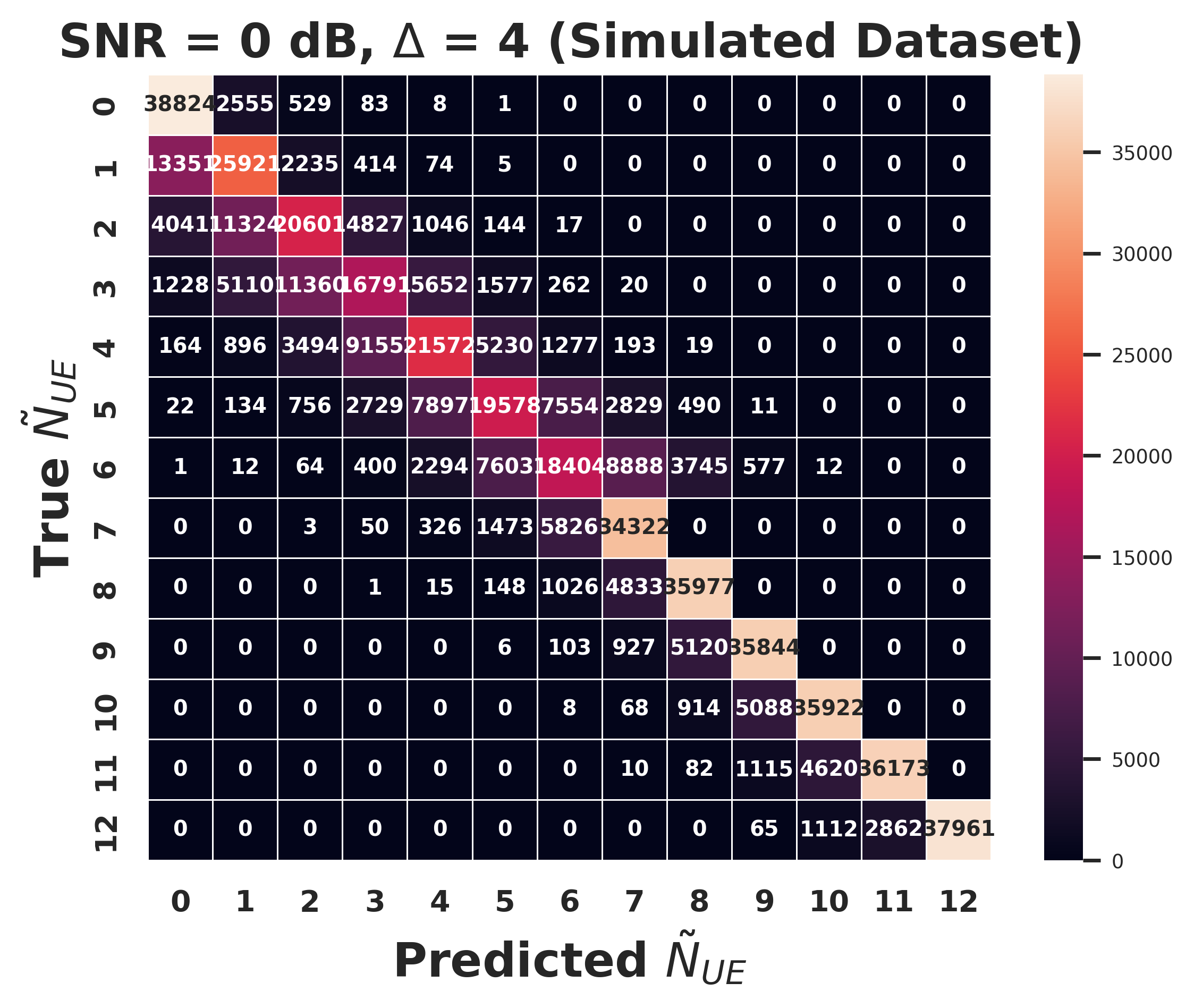}

    \caption{{Confusion matrix for the number of multiplexed UEs $\Tilde{N}_{UE}$, at SNR = $0$ dB and $\Delta$ = $4$ for simulated test dataset. }}
    \label{fig: cm_num_UE_snr_0_off_4}
\end{figure}

\begin{figure}[ht!]
    \captionsetup{justification=justified}

     \centering
     \includegraphics[width=0.48\textwidth]{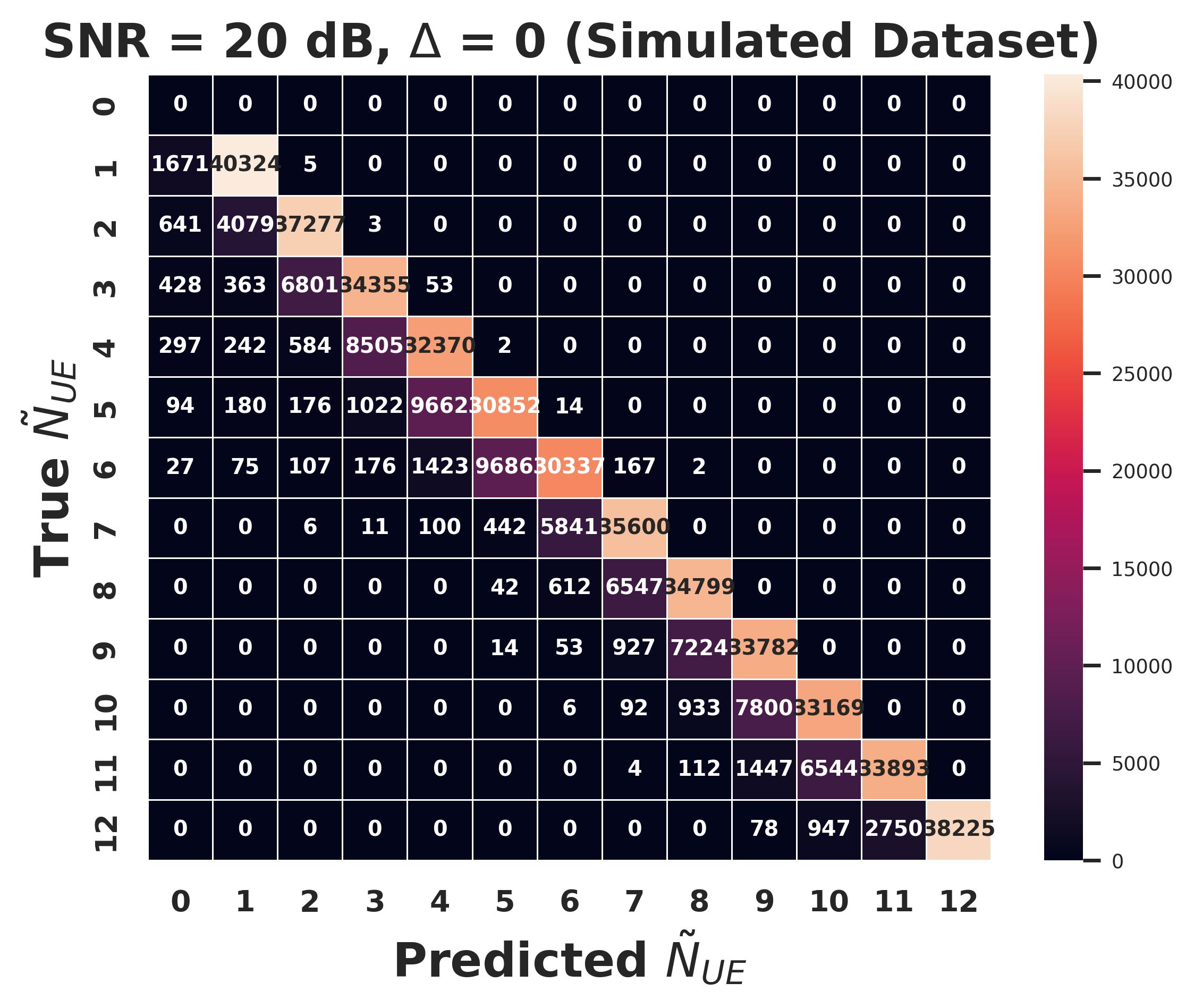}

    \caption{{Confusion matrix for the number of multiplexed UEs $\Tilde{N}_{UE}$, at SNR = $20$ dB and $\Delta$ = $0$ for simulated test dataset.}}
    \label{fig: cm_num_UE_snr_20_off_0}
\end{figure}

\begin{figure}[ht!]
    \captionsetup{justification=justified}

     \centering
     \includegraphics[width=0.48\textwidth]{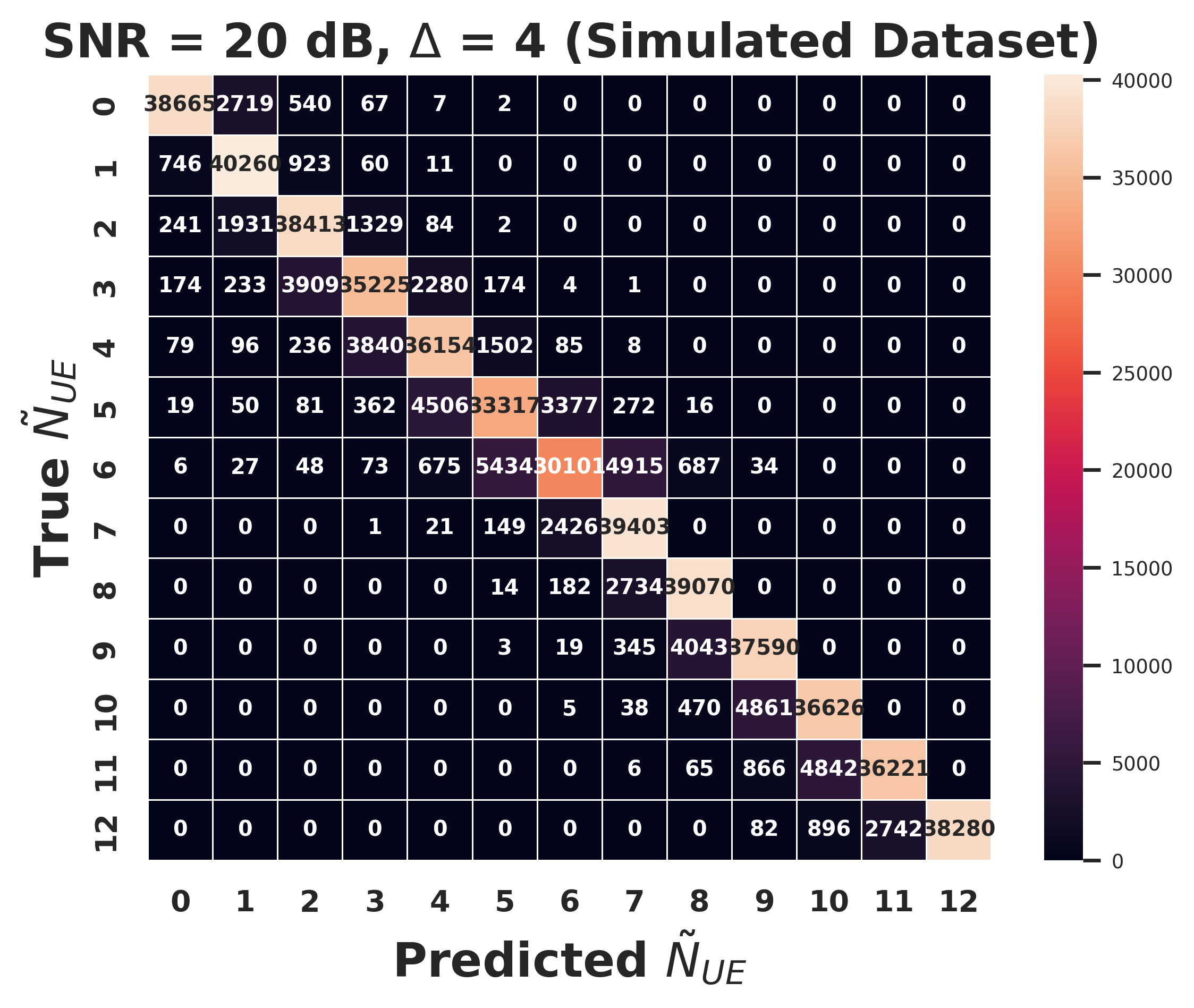}

    \caption{{Confusion matrix for the number of multiplexed UEs $\Tilde{N}_{UE}$, at SNR = $20$ dB and $\Delta$ = $4$ for simulated test dataset.}}
    \label{fig: cm_num_UE_snr_20_off_4}
\end{figure}

It should be noted that the multi-label confusion matrix does not show the exact details of misclassifications. To gain a deeper understanding of the mistakes made by the model, consider the matrices in Figures~\ref{fig: cm_num_UE_snr_0_off_0} to~\ref{fig: cm_num_UE_snr_20_off_4}. For each instance of the received PUCCH signal, there are two characteristics of correct classification. The first characteristic is correctly identifying the number of multiplexed UEs ($\Tilde{N}_{UE}$) embedded in the signal. Even though this is not an explicit output of the NN model, we believe that the model is learning to estimate the number of multiplexed UEs. Our belief is based on the fact that when the number of possible multiplexed UEs is not given as metadata to the NN input, the accuracy drops significantly. The second characteristic of correct classification is derived from the explicit output of the NN model, which is the correct prediction of the value of $\alpha$ for each of the multiplexed UEs. 

Figures~\ref{fig: cm_num_UE_snr_0_off_0} and ~\ref{fig: cm_num_UE_snr_0_off_4} show confusion matrices for the number of multiplexed UEs for simulated data at an SNR of $0$ dB and $\Delta$ values of $0$ and $4$, respectively. Figures~\ref{fig: cm_num_UE_snr_20_off_0} and ~\ref{fig: cm_num_UE_snr_20_off_4} are confusion matrices for the number of multiplexed UEs for simulated data at an SNR of $20$ dB and $\Delta$ values of $0$ and $4$, respectively. We notice that all the confusion matrices for the prediction of the number of multiplexed UEs are largely diagonally dominant. Furthermore, for any SNR, when $\Delta = 0$, the error in the predicted number of UEs fluctuates between +1 and -1 in the majority of instances, confirming the high accuracy values in Figure~\ref{fig: acc_vs_snr}. As $\Delta$ increases,  the error tends to grow on the positive side, meaning that the NN model predicts more UEs than are actually present. As stated above, the overprediction is because $\Delta$ is always positive. 

\begin{figure}[ht!]
    \captionsetup{justification=justified}

     \centering
     \includegraphics[width=0.48\textwidth]{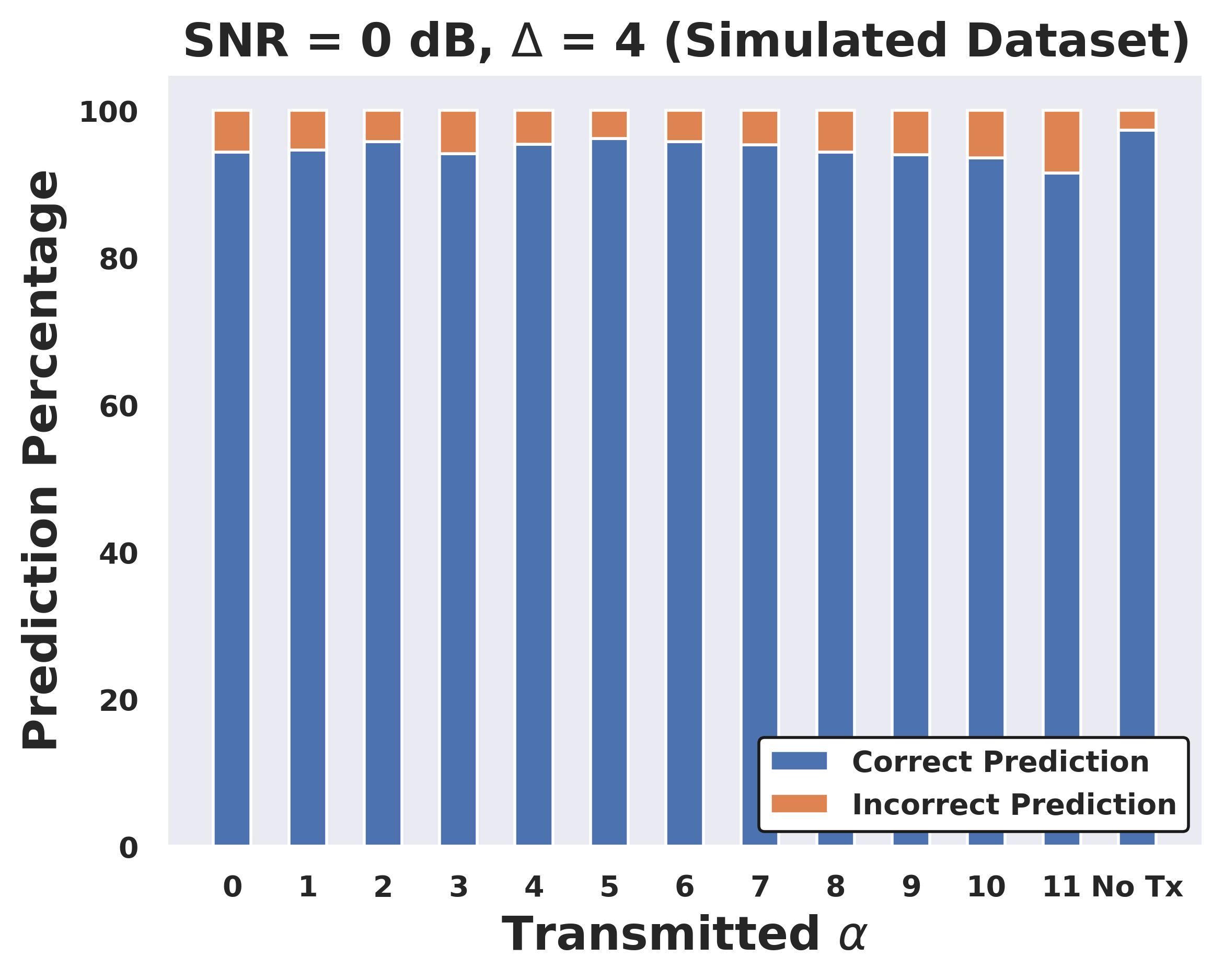}

    \caption{{Column chart for the cyclic shift $\alpha$ at SNR = $0$ dB, $\Delta$ = $4$, and with $\Tilde{N}_{UE} = \{0, 1, 2, \dots, 12\}$ for simulated test dataset.}}
    \label{fig: bar_alpha_SNR_snr_0_off_4}
\end{figure}

\begin{figure}[ht!]
    \captionsetup{justification=justified}

     \centering
     \includegraphics[width=0.48\textwidth]{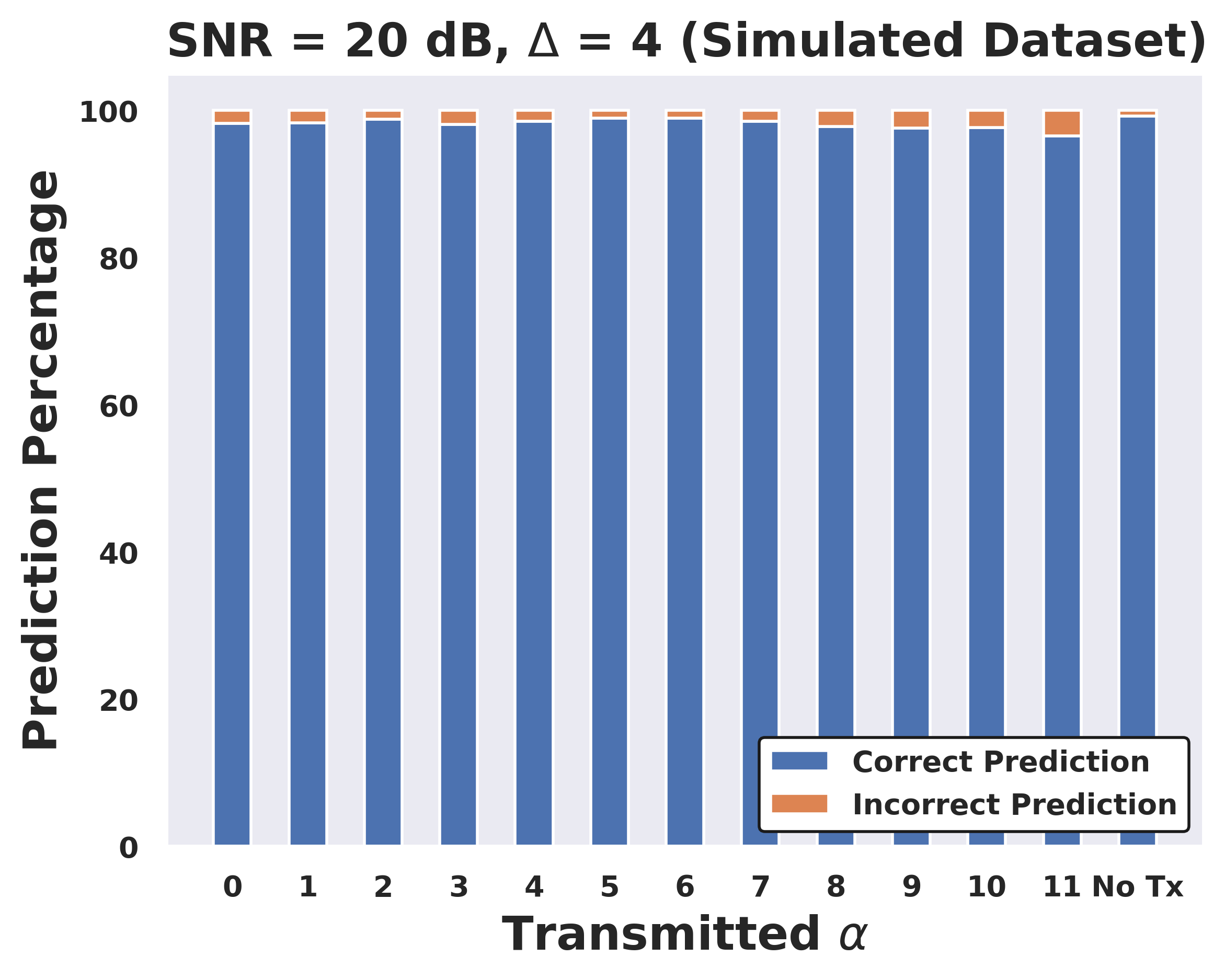}

    \caption{{Column chart for the cyclic shift $\alpha$ at SNR = $20$ dB, $\Delta$ = $4$, and with $\Tilde{N}_{UE} = \{0, 1, 2, \dots, 12\}$ for simulated test dataset.}}
    \label{fig: bar_alpha_SNR_snr_20_off_4}
\end{figure}

The column charts in Figure~\ref{fig: bar_alpha_SNR_snr_0_off_4} and Figure~\ref{fig: bar_alpha_SNR_snr_20_off_4} show the percentage of correct and incorrect predictions for the transmission of every $\alpha$ from 0 through 11. The result of the NN model's prediction for the scenario where a particular $\alpha$ has not been transmitted by any UE is also captured in the last column labeled $No\ Tx$. Correct prediction in the last column indicates that the model was able to correctly identify the non-transmission of a particular $\alpha$ value, and incorrect detection determines the instances where a particular $\alpha$ has not been transmitted but incorrectly predicted. 

From Figure~\ref{fig: bar_alpha_SNR_snr_0_off_4} and Figure~\ref{fig: bar_alpha_SNR_snr_20_off_4}, it is evident that the NN model is not biased towards any particular $\alpha$ and the cyclic shifts are correctly identified in the majority of instances.


\subsection{IMPLICATIONS OF MODEL PERFORMANCE ON A 5G SYSTEM}
In general, both false detections and missed detections are detrimental to PUCCH Format 0 performance. However, from the context of sustaining a communication link between a gNB and a UE, a small number of false detections are more tolerable than missed detections. For example, when a gNB falsely detects an SR, it allocates a small uplink bandwidth to the UE. Since the UE does not have any valid data to transmit, it would populate the uplink packet with padding information and indicate its lack of data in the Buffer Status Report (BSR)~\cite{Lin_2018}. No further uplink grants will be allocated to the UE. 

On the other hand, if a gNB misses the detection of an ACK, it will assume that the prior downlink transmission was not decoded by the UE. In this scenario, a significant downlink bandwidth will be allocated to the UE for the retransmission. Successive missed detection of ACKs will lead to the termination of the radio link (referred to as Radio Link Failure) between the gNB and the UE. 

Hence, occupying a small uplink bandwidth (due to false detection of an SR) is more tolerable compared to occupying a large downlink bandwidth (due to missed detection of ACKs), since recovery from a Radio Link Failure (RLF) involves heavy signaling for the UE to reattach. 

As shown in Figures~\ref{fig: cm_multilabel_snr_20_off_4}, ~\ref{fig: cm_num_UE_snr_0_off_4}, and ~\ref{fig: cm_num_UE_snr_20_off_4}, at low SNRs or high $\Delta$ values, the proposed UCINet0 tends to overpredict the number of UEs actually present, rather than underpredict. This shows that the model, even when it makes a wrong prediction, is good at avoiding scenarios leading to RLFs.

\subsection{MODEL COMPLEXITY, MEMORY AND LATENCY ANALYSIS}

\begin{figure}[h]
\centering
\includegraphics[width=0.48\textwidth]{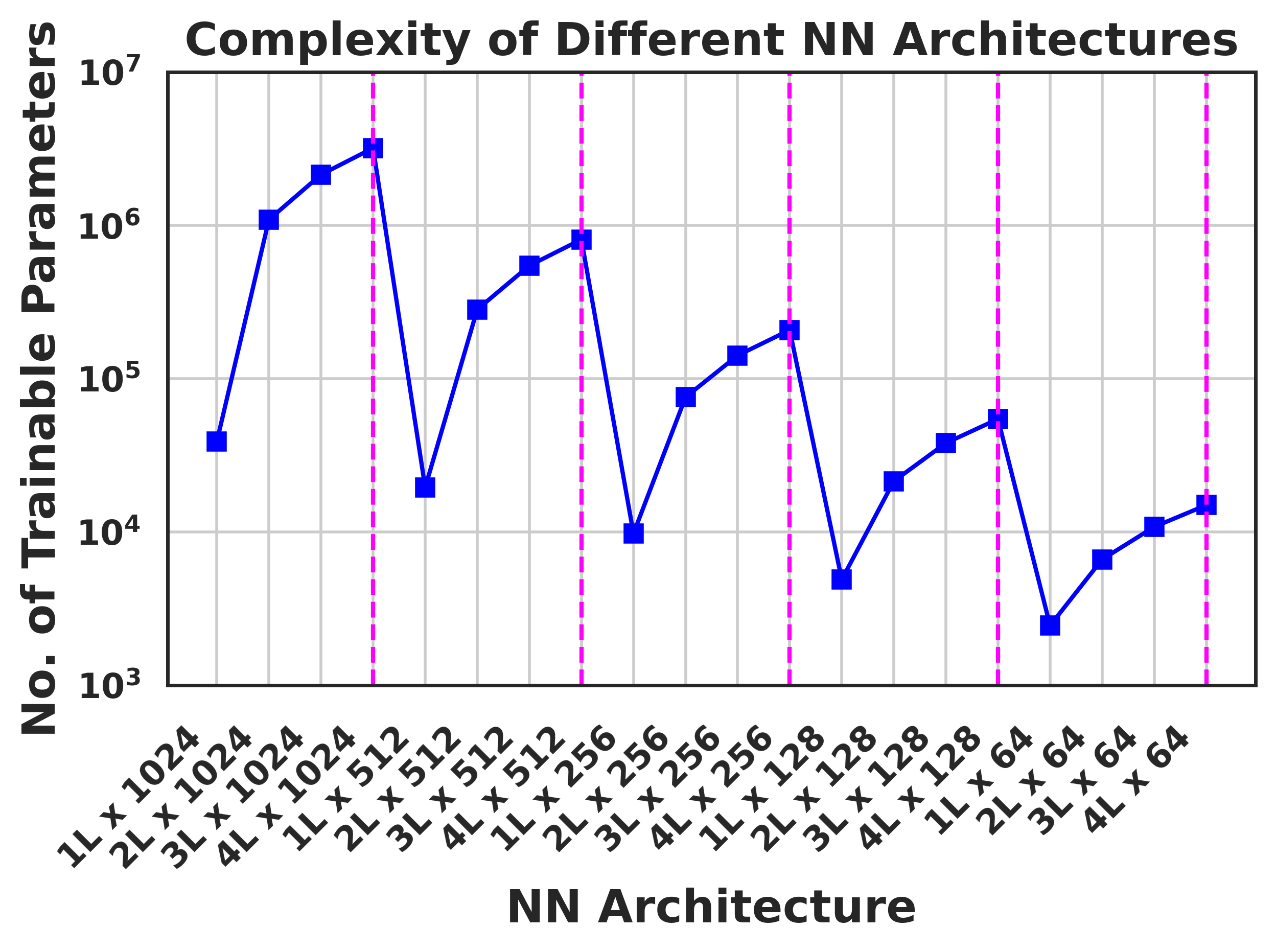}
\caption{{Line graph showing the number of trainable parameters for different FCN architectures. The number of FCN layers and the number of neurons of each architecture are presented on the x-axis. The number of trainable parameters corresponding to these architectures increases with the number of layers and neurons.}}
\label{fig: weights_complexity}
\end{figure}

Figure~\ref{fig: weights_complexity} shows the number of trainable parameters for different FCN architectures. The number of trainable parameters increases with the number of layers and neurons. An increase in the number of trainable parameters leads directly to an increase in the latency due to (1) fetching the weights from memory and (2) corresponding neural network computations. Current 5G systems function with fundamental time units in the order of hundreds of microseconds. To cater to this requirement, processing all the physical layer channels allocated to each user must be completed in a duration of tens of microseconds. For example,  for a subcarrier spacing of $30$kHz, the slot duration is of $500\mu s$. Hence, the physical layer processing of all the users allocated in that slot needs to be completed within this time frame. This time constraint also applies to AI/ML-based approaches such as neural networks. Hence, FPGAs are a promising choice for hardware implementation of such time-constrained neural networks.

In order to deploy a version of UCINet0 on hardware devices such as FPGAs, the trade-offs between model performance (Figure~\ref{fig: acc_vs_complexity}) and model complexity (Figure~\ref{fig: weights_complexity}) must be considered. These two parameters help us choose the best-suited model based on the available memory and compute resources on the hardware. {For the inference of UCINet0 for various test scenarios, we use an NVIDIA A100 Tensor Core GPU. Recall that our NN consists of an input layer with 24 neurons, two hidden layers with 257 and 256 neurons respectively, and an output layer with 12 neurons. This model consists of $75532$ trainable parameters ($(24\times256+256) +( 257\times256+256)+(256\times12+12)$). Similarly, the number of trainable parameters is calculated for other architectures and shown in Figure~\ref{fig: weights_complexity}.}

{Assuming each parameter is represented by single precision floating point format (FP32), i.e., 4 bytes per parameter, resulting in a model memory of $302.128$KB. This is the memory required to store the model's weights and biases both during training and inference. Along with this, activation memory is used to store the outputs of each layer during the forward pass of the network. With a batch size of 512, the total activation memory consumed by the model is $512 \times (25 + 256 +257 +256 +12) \times 4 \text{bytes} = 1.65$MB. To perform an inference on 1000 instances on the A100, the computed average runtime latency per instance is $0.27$ ms which is sufficient for a 5G network with $30$KHz which has a slot duration of $0.5$ ms.}

The memory and bandwidth requirements for storing and retrieving the weights of neural networks like UCINet0 can be met by Xilinx's latest Versal series FPGAs~\cite{wierse2023evaluation}, which contain AI engines to perform matrix multiplications efficiently.

\section*{CONCLUSION}

In this paper, we have designed a generalized Machine Learning based receiver for PUCCH Format 0 that decodes all the combinations of UCI bits across several multiplexed users. The proposed UCINet0 model is a multi-label classifier that demonstrates superior performance under both slow and fast fading channel scenarios, compared to conventional correlation-based approaches by a good margin. The inclusion of hardware-captured datasets in the testing of the model adds robustness and shows the ability of the network to work in a wide range of scenarios. In addition to standard AI/ML metrics like accuracy, we have also provided insights into the interpretability of the NN model through confusion matrices and column charts of detection percentages of individual classes.

\section*{ACKNOWLEDGMENT}
The authors would like to thank the Department of Telecommunications (DOT), India for funding the 5G Testbed project and the Ministry of Electronics and Information Technology (MeitY) for funding this work through the project ``Next Generation Wireless Research and Standardization on 5G and Beyond".


\bibliographystyle{IEEEtran}
\bibliography{bibfile}

\newpage

\vfill\pagebreak

\end{document}